\documentclass[11pt,a4paper]{article}
\usepackage[dvipsnames]{xcolor}
\usepackage{jheppub}
\usepackage{amssymb}
\usepackage{amsfonts}
\usepackage{amsthm}
\usepackage{amsmath}
\usepackage{mathrsfs}
\usepackage{footnote}
\usepackage{graphicx} 
\usepackage{subcaption}
\usepackage{stmaryrd}
\usepackage{cleveref}
\usepackage{hyperref}
\usepackage{tcolorbox}
\usepackage{tikz}  
\usepackage{tikzsymbols}
\usepackage{physics}
\usepackage{microtype}
\usepackage{bbm}
\usepackage{mathtools}
\usepackage{enumitem}
\usepackage{nicefrac}
\usepackage{leftindex}
\usepackage{footmisc}
\usepackage{booktabs}
\usepackage{centernot}
\usepackage{stmaryrd}
\usepackage{enumitem}
\usepackage{footnote}
\usepackage{twemojis}
\usepackage{mathptmx}
\usepackage{anyfontsize}
\usepackage{t1enc}
\hypersetup{colorlinks=true, linkcolor=Blue, citecolor=Blue, urlcolor=Blue}
\usepackage[a4paper, left=6cm, right=1cm, top=4.98cm, bottom=1.98cm]{geometry}

\title{The firewall paradox is Wigner's friend paradox}

\author{Ladina Hausmann and}
\author{Renato Renner}
\affiliation{\small \it Institute for Theoretical Physics, ETH Z\"urich}
\emailAdd{hladina@ethz.ch}
\emailAdd{renner@ethz.ch}

\notoc

\makesavenoteenv{itemize}
\makesavenoteenv{figure}
\makesavenoteenv{enumerate}

\definecolor{shadecolor}{gray}{.85}%
\definecolor{tintedcolor}{gray}{.80}%

\setlist{nosep}

\makeatletter
\newcommand{\xMapsto}[2][]{\ext@arrow 0599{\Mapstofill@}{#1}{#2}}
\def\Mapstofill@{\arrowfill@{\Mapstochar\Relbar}\Relbar\Rightarrow}
\makeatother

\usetikzlibrary{quantikz2}
\usetikzlibrary{decorations.pathmorphing}
\usetikzlibrary{arrows.meta}
\usetikzlibrary{decorations.markings}
\tikzset{wavy/.style={decorate, decoration=snake}}

\newtheorem{theorem}{Theorem}

\definecolor{tintedcolor}{gray}{.85}%

\newtheorem{innercustomgeneric}{\customgenericname}
\providecommand{\customgenericname}{}
\newcommand{\newcustomtheorem}[3]{%
  \newenvironment{#1}[1]
  {%
   \ifdefined\crefalias\crefalias{innercustomgeneric}{#2}\fi
   \renewcommand\customgenericname{#2}%
   \renewcommand\theinnercustomgeneric{##1}%
   \innercustomgeneric
  }
  {\endinnercustomgeneric}%
  \ifdefined\crefname\crefname{#2}{#3}{#3s}\fi
}

\newcustomtheorem{customAssumption}{Assumption}{assumption}
\newcustomtheorem{customPrinciple}{Principle}{principle}

\makeatletter
\newcommand{\myitem}[1]{%
\item[\textbf{#1:}]\protected@edef\@currentlabel{#1}%
}
\makeatother

\theoremstyle{definition}
\newtheorem{protocol}{Protocol}

\crefname{protocol}{protocol}{protocols}

\newtheoremstyle{named}{}{}{}{}{\bfseries}{.}{.5em}{\thmnote{#3 }#1}
\theoremstyle{named}
\newtheorem*{namedgame}{Game}

\makeatletter
\def\namedlabel#1#2{\begingroup
   \def\@currentlabel{#2}%
   \label{#1}\endgroup
}
\makeatother

\makeatletter
\pgfkeys{/pgf/document dog ear/.value required}%
\pgfkeys{/pgf/document dog ear/.code={%
        \pgfmathparse{#1}%
        \edef\pgfmathresult{\pgfmathresult pt}%
        \pgfkeyslet{/pgf/document dog ear}{\pgfmathresult}%
    }%
}%
\pgfkeys{/pgf/document dog ear=7.5pt}%

\pgfdeclareshape{document}{
    \inheritsavedanchors[from=rectangle] 
    \inheritanchorborder[from=rectangle]
    \inheritanchor[from=rectangle]{center}
    \inheritanchor[from=rectangle]{north}
    \inheritanchor[from=rectangle]{south}
    \inheritanchor[from=rectangle]{west}
    \inheritanchor[from=rectangle]{east}
    \backgroundpath{
        \southwest \pgf@xa=\pgf@x \pgf@ya=\pgf@y
        \northeast \pgf@xb=\pgf@x \pgf@yb=\pgf@y
        \pgf@xc=\pgf@xb \advance\pgf@xc by-\pgfkeysvalueof{/pgf/document dog ear}
        \pgf@yc=\pgf@yb \advance\pgf@yc by-\pgfkeysvalueof{/pgf/document dog ear}
        \pgfpathmoveto{\pgfpoint{\pgf@xa}{\pgf@ya}}
        \pgfpathlineto{\pgfpoint{\pgf@xa}{\pgf@yb}}
        \pgfpathlineto{\pgfpoint{\pgf@xc}{\pgf@yb}}
        \pgfpathlineto{\pgfpoint{\pgf@xb}{\pgf@yc}}
        \pgfpathlineto{\pgfpoint{\pgf@xb}{\pgf@ya}}
        \pgfpathclose
        \pgfpathmoveto{\pgfpoint{\pgf@xc}{\pgf@yb}}
        \pgfpathlineto{\pgfpoint{\pgf@xc}{\pgf@yc}}
        \pgfpathlineto{\pgfpoint{\pgf@xb}{\pgf@yc}}
        \pgfpathlineto{\pgfpoint{\pgf@xc}{\pgf@yc}}
    }%
}%
\makeatother

\newcommand{\qubit}{
    \begin{tikzpicture}[scale=0.4]
        \fill[black] (0,0) circle (0.2cm);
        \draw[-stealth, very thick] (-0.2,-0.4) -- (0.3,0.6); 
    \end{tikzpicture}
}

\newcommand{\meterComp}[1]{
  \begin{tikzpicture}
  \node[#1, scale = 0.7] at  (0,0) {
      \begin{quantikz}[background color=tintedcolor]
        \meter[1]{}
      \end{quantikz}
  };
  \node[#1, scale = 0.7] at (-0.6,-0.1) {$\mathrm{comp.}$};
  \end{tikzpicture}
}

\newcommand{\friendCircuit}{
  \begin{tikzpicture}
    \node[anchor=center,MidnightBlue, scale=0.7,rotate = 90] at (0,0) {%
    \begin{quantikz}[background color=tintedcolor]
      \lstick{\text{\rotatebox{270}{$\ket{\bar{0}}$} }} &  \wire[l][1]["\text{\rotatebox{270}{$Q$}}"{above,pos=0.75}]{a} & &
    \end{quantikz}
    };

    \node[] at (-0.25,0.25) {\meterComp{MidnightBlue}};
  \end{tikzpicture}
}

\newcommand{\friendDeutschCircuit}{
  \begin{tikzpicture}
    \node[anchor=center,MidnightBlue, scale=0.7,rotate = 90] at (0,0) {%
    \begin{quantikz}[background color=tintedcolor]
      &  \wire[l][1]["\text{\rotatebox{270}{$Q$}}"{above,pos=0.75}]{a} & \setwiretype{c}& \wire[l][1]["\text{\rotatebox{270}{$R_A$}}"{above,pos=0.75}]{a}
    \end{quantikz}
    };

    \node[] at (-0.2,0) {\meterComp{MidnightBlue}};
  \end{tikzpicture}
}

\newcommand{\bobDeutschCircuit}{
  \begin{tikzpicture}
    \node[anchor=center,BrickRed, scale=0.7, rotate = 90] at (0,0) {%
      \begin{quantikz}[background color=tintedcolor]
        & 
        \gate[2]{\text{\rotatebox{270}{$U^{\phantom{\dagger}}$}}} \wire[l][1]["\text{\rotatebox{270}{$Q$}}"{above,pos=0.75}]{a}&\gate[2]{\text{\rotatebox{270}{$U^\dagger$}}}&  &\setwiretype{c}&\wire[l][1]["\text{\rotatebox{270}{$R_B$}}"{above,pos=0.9}]{a}\\
        &\wire[l][1]["\text{\rotatebox{270}{$L_A$}}"{above,pos=0.75}]{a} && 
      \end{quantikz}
    };
    \node[] at (-0.65,0.75) {\meterDiag};
  \end{tikzpicture}
}

\newcommand{\wignerCircuit}{
  \begin{tikzpicture}
    \node[anchor=center,BrickRed, scale=0.7, rotate = 90] at (-3.2,-2) {%
      \begin{quantikz}[background color=tintedcolor]
        \lstick{\text{\rotatebox{270}{$\ket{\bar{0}}$} }} & \gate[2]{\text{\rotatebox{270}{$U$}}}&\wire[l][1]["\text{\rotatebox{270}{$Q$}}"{above,pos=2.8}]{a} \\
         \lstick{\text{\rotatebox{270}{$\ket{\varphi}$} }}&\wire[l][1]["\text{\rotatebox{270}{$L_A$}}"{above,pos=0.75}]{a} & 
      \end{quantikz}
    };
  \end{tikzpicture}
}

\newcommand{\AliceFR}[1]{
  \begin{tikzpicture}
    \node[anchor=center,#1, scale=0.7, rotate = 90] at (-3.2,-2) {%
      \begin{quantikz}[background color=tintedcolor]
        \lstick[2]{\text{\rotatebox{270}{$\psi$} }} &\wire[l][1]["\text{\rotatebox{270}{$Q_A$}}"{above,pos=0.85}]{a}  &\setwiretype{c}\wire[l][1]["\text{\rotatebox{270}{$R_A$}}"{above,pos=0.7}]{a} \\
         &\gate[2]{\text{\rotatebox{270}{$U_{\mathrm{Charly}}^{\phantom{\dagger}}$}}}&\gate[2]{\text{\rotatebox{270}{$U_{\mathrm{Charly}}^{\dagger}$}}} & \wire[l][1]["\text{\rotatebox{270}{$Q_C$}}"{above,pos=5.6}]{a}\\
        &\wire[l][1]["\text{\rotatebox{270}{$L_C$}}"{above,pos=0.8}]{a} & &
      \end{quantikz}
    };
  \end{tikzpicture}
}

\newcommand{\AliceCircuitFR}{
  \begin{tikzpicture}
    \node[] at (0,0){\AliceFR{MidnightBlue}};
    \node[] at (-1.4,-0.2){\meterComp{MidnightBlue}};
  \end{tikzpicture}
}

\newcommand{\CharlyFR}[1]{
  \begin{tikzpicture}
    \node[anchor=center,#1, scale=0.7, rotate = 90] at (-3.2,-2) {%
      \begin{quantikz}[background color=tintedcolor]
        \lstick[2]{\text{\rotatebox{270}{$\psi$} }} &\wire[l][1]["\text{\rotatebox{270}{$Q_A$}}"{above,pos=0.75}]{a} \\
        &\wire[l][1]["\text{\rotatebox{270}{$Q_C$}}"{above,pos=0.8}]{a}&&\setwiretype{c}\wire[l][1]["\text{\rotatebox{270}{$R_C$}}"{above,pos=0.65}]{a}
      \end{quantikz}
    };
  \end{tikzpicture}
}

\newcommand{\CharlyCircuitFR}{
  \begin{tikzpicture}
    \node[] at (0,0){\CharlyFR{Violet}};
    \node[Violet, scale = 0.7] at  (0.3,0.27) {
      \begin{quantikz}[background color=tintedcolor]
        \meter[1]{}
      \end{quantikz}
  };
  \node[Violet, scale = 0.7] at (-0.3,0.2) {$\mathrm{comp.}$};
  \end{tikzpicture}
}

\newcommand{\BobCircuitFR}{
  \begin{tikzpicture}
    \node[anchor=center,BrickRed, scale=0.7, rotate = 90] at (0,0) {%
    \begin{quantikz}[background color=tintedcolor]
      &\wire[l][1]["\text{\rotatebox{270}{$L_A$}}"{above,pos=0.75}]{a}&  \gate[2]{\text{\rotatebox{270}{$U_{\mathrm{Alice}}^{\phantom{\dagger}}$}}}&\gate[2]{\text{\rotatebox{270}{$U_{\mathrm{Alice}}^{\dagger}$}}}&   \\
      \lstick[2]{\text{\rotatebox{270}{$\psi$}}}  &\wire[l][1]["\text{\rotatebox{270}{$Q_A$}}"{above,pos=0.75}]{a}&&& &&\setwiretype{c} \wire[l][1]["\text{\rotatebox{270}{$R_B$}}"{above,pos=0.65}]{a} \\
      & \wire[l][1]["\text{\rotatebox{270}{$Q_C$}}"{above,pos=0.75}]{a}&&&\\
    \end{quantikz}
    };
    \node[] at (-0.1,1.32){\meterDiag};
  \end{tikzpicture}
}

\newcommand{\RefCircuitFR}{
  \begin{tikzpicture}
    \node[anchor=center,RedOrange, scale=0.7,rotate = 90] at (0,0) {%
    \begin{quantikz}[background color=tintedcolor]
      &\wire[l][1]["\text{\rotatebox{270}{$L_A$}}"{above,pos=0.75}]{a}&  \gate[2]{\text{\rotatebox{270}{$U_{\mathrm{Alice}}^{\phantom{\dagger}}$}}}&\gate[2]{\text{\rotatebox{270}{$U_{\mathrm{Alice}}^{\dagger}$}}}& \\
      \lstick[2]{\text{\rotatebox{270}{$\psi$}}}  &\wire[l][1]["\text{\rotatebox{270}{$Q_A$}}"{above,pos=0.75}]{a}& && \\
      &\wire[l][1]["\text{\rotatebox{270}{$Q_C$}}"{above,pos=0.75}]{a}&&\gate[2]{\text{\rotatebox{270}{$U_{\mathrm{Charly}}^{\phantom{\dagger}}$}}}&\gate[2]{\text{\rotatebox{270}{$U_{\mathrm{Charly}}^{\dagger}$}}}&\wire[l][1]["\text{\rotatebox{270}{$Q_C$}}"{above,pos=-0.75}]{a}&\\
      &\wire[l][1]["\text{\rotatebox{270}{$L_C$}}"{above,pos=0.75}]{a} &&&& \\
    \end{quantikz}
    };

  \end{tikzpicture}
}
\newcommand{\HaydenPreskillPenrose}{
  \begin{tikzpicture}
    \draw[black, thick] (0,6)--(0,0)  --  (4.5,4.5) ;
    \draw[black, thick] (4.5,4.5)--(3,6) ;
    \draw[black, wavy] (3,6) -- (0,6);
    \draw[black, dotted] (3,6) -- (0,3);

    \draw[MidnightBlue, thick] plot [smooth] coordinates {(0,0) (1,2) (1.6,5) (2,6)};
    \node[MidnightBlue] at (1.4,5.2) {$Q$};

    \draw[BrickRed, thick] plot [smooth] coordinates {(0,0) (1.2,2) (2.5,4) (3,6)};
    \node[BrickRed] at (3.05,5.05) {$Q_R$};
    \node[BrickRed] at (1.25,4.05) {$Q$};

    \draw[ForestGreen, thick, dashed] plot [smooth] coordinates {(0,4.9)(2,5)(4.5,4.5)};

    \node[Plum] at (2.2,4.6) {$R'$}; 
    \draw[-latex, Plum, decorate, decoration={snake, amplitude=0.02cm, segment length=0.1cm}]
    (1.35,3.55)--(2.8,4.9);
    \node[Plum] at (2.8,4.2) {$R$};   
    \draw[-latex, Plum, decorate, decoration={snake, amplitude=0.02cm, segment length=0.1cm}] 
    (1.7,3.3)--(2.6,4.2);
  \end{tikzpicture}
}

\newcommand{\HaydenPreskillAlice}{
    \begin{tikzpicture}
       \node[MidnightBlue, scale = 0.7, rotate = 90] at (5,1.2){
          \begin{quantikz}[background color=tintedcolor]
            \lstick{\text{\rotatebox{270}{\phantom{$\psi$}}}}&  
            \wire[l][1]["\text{\rotatebox{270}{$Q$}}"{above,pos=0.75}]{a}  \slice[style =
            ForestGreen]{}  && &\setwiretype{c}\wire[l][1]["\text{\rotatebox{270}{$R_A$}}"{above,pos=0.65}]{a}
          \end{quantikz}
        };
        \node[] at (4.8,1.58) {\meterComp{MidnightBlue}};  
    \end{tikzpicture}       
}

\newcommand{\HaydenPreskillBob}{
    \begin{tikzpicture}
        \node[BrickRed, scale = 0.7, rotate = 90] at (4,-1.8){
          \begin{quantikz}[background color=tintedcolor]
            & \gate[2]{\text{\rotatebox{270}{$U_{\mathrm{Evap}}$}}}&&\slice[style =
            ForestGreen]{}& \wire[l][1]["\text{\rotatebox{270}{$B \ H$}}"{above,pos=6.1}]{a}&\\
            \lstick{\text{\rotatebox{270}{\phantom{$\psi$}} }} & &\gate[2]{\text{\rotatebox{270}{$U_{\mathrm{Recov}}$}}}& \wire[l][1]["\text{\rotatebox{270}{$Q$} }"{above,pos=5}]{a}& \wire[l][1]["\text{\rotatebox{270}{$R'$} }"{above,pos=3.7}]{a}&\\
            & && \wire[l][1]["\text{\rotatebox{270}{$R$}}"{above,pos=5.1},"\text{\rotatebox{270}{$Q_R$}}"{above,pos=0.65}]{a}&&&\setwiretype{c}\wire[l][1]["\text{\rotatebox{270}{$R_B$}}"{above,pos=0.55}]{a}
          \end{quantikz}
        };
        \node[] at (4.9,-0.65) {\meterDiag};
    \end{tikzpicture}     
}

\newcommand{\firewallPenrose}{
  \begin{tikzpicture}
      \draw[black, thick] (0,6)--(0,0)  --  (4.5,4.5) ;
      \draw[black, thick] (4.5,4.5)--(3,6) ;
      \draw[black, wavy] (3,6) -- (0,6);
      \draw[black, dotted, thick] (3,6) -- (0,3);

      \node[Plum] at (1.3,4.78) {$A$};
      \draw[-latex, Plum, decorate, decoration={snake, amplitude=0.02cm, segment length=0.1cm}]
      (0.9,4.1)--(1.5,4.7);
      \node[Plum] at (1.85,4.4) {$B$};   
      \draw[-latex, Plum, decorate, decoration={snake, amplitude=0.02cm, segment length=0.1cm}] (1.1,3.9)--(1.65,4.45);

      \draw[BrickRed, double] (2.4,4) -- (1.5,5.49);
      \draw[tintedcolor,line width=.589pt] (2.4,4) -- (1.5,5.5);
      \draw[BrickRed, thick] plot [smooth] coordinates {(0,0) (2,3)(3,6)};
      \node[BrickRed] at (2.6,4) {$R$};
      \node[BrickRed] at (2.2,5.05) {$R_{\mathrm{Bob}}$};

      \draw[gray, thick] plot [smooth] coordinates {(2,3) (1.6,4.5)  (1.5,5.5) (1.5,6)};

      \draw[MidnightBlue, thick] plot [smooth] coordinates {(2,3) (1.5,4.5) (1.5,5.5) (1.5,6)};
  \end{tikzpicture}
}

\newcommand{\AliceCircuit}{
  \begin{tikzpicture}
    \node[MidnightBlue, scale = 0.7, rotate = 90] at  (7,7.5) {
      \begin{quantikz}[background color=tintedcolor]
        & \gate[1]{\text{\rotatebox{270}{$V_{\mathrm{Alice}}$}}}&&\slice[style = MidnightBlue, label
        style=MidnightBlue]{\text{\rotatebox{270}{$\psi^{\mathrm{Alice}}$}}} &
        \wire[l][1]["\text{\rotatebox{270}{$Q_\mathrm{Alice}$}}"{above,pos=2.5},"\text{\rotatebox{270}{$A$}}"{above,pos=4.9}]{a} & \setwiretype{c} \wire[l][1]["\text{\rotatebox{270}{$R_\mathrm{Alice}$}}"{above,pos=-0.22}]{a} &\\
        &\gate[1]{\text{\rotatebox{270}{\phantom{.}$V_{\mathrm{Ref}}$\phantom{.}}}}&&& \wire[l][1]["\text{\rotatebox{270}{$Q_\mathrm{Ref}$}}"{above,pos=2.5},"\text{\rotatebox{270}{$B$}}"{above,pos=4.9}]{a}
      \end{quantikz}
    };
    \node[MidnightBlue] at (6.5,8.25) {\meterComp{MidnightBlue}};
  \end{tikzpicture}
}

\newcommand{\firewallAlice}{
  \begin{tikzpicture}
    \fill[MidnightBlue!30] (7,5) circle (1.5);
    \draw[color=black, dotted, thick](7,5) circle (0.75);
    \node[MidnightBlue] at (6.8,5) {$A$};
    \node[MidnightBlue] at (5.9,5) {$B$};
    \draw[thick, decorate, decoration={snake, amplitude=0.04cm, segment length=0.2cm}] 
    (6.7,5) -- (6,5);
    \node[] at (3.5,5) {\AliceCircuit};
    
  \end{tikzpicture}
}

\newcommand{\meterDiag}{
  \begin{tikzpicture}
  \node[BrickRed, scale = 0.7] at  (0,0) {
      \begin{quantikz}[background color=tintedcolor]
        \meter[1]{} 
      \end{quantikz}
  };
  \node[BrickRed, scale = 0.7] at (-0.55,-0.05) {$\mathrm{diag.}$};
  \end{tikzpicture}
}

\newcommand{\BobCircuit}{
  \begin{tikzpicture}
    \node[BrickRed, scale = 0.7,rotate = 90] at  (6,-2) {
        \begin{quantikz}[background color=tintedcolor]
          & \gate[1]{\text{\rotatebox{270}{\phantom{.}$V_{\mathrm{Bob}}$\phantom{.}}}}&&\slice[style = BrickRed, label style=BrickRed]{\text{\rotatebox{270}{$\psi^{\mathrm{Bob}}$}}}&& \setwiretype{c} & \wire[l][1]["\text{\rotatebox{270}{$Q_\mathrm{Bob}$}}"{above,pos=4.6},"\text{\rotatebox{270}{$R$}}"{above,pos=6.9},
          "\text{\rotatebox{270}{$R_\mathrm{Bob}$}}"{above,pos=0.7}]{a}\\
          &\gate[1]{\text{\rotatebox{270}{\phantom{.}$V_{\mathrm{Ref}}$\phantom{.}}}}&&& \wire[l][1]["\text{\rotatebox{270}{$Q_\mathrm{Ref}$}}"{above,pos=2.65},"\text{\rotatebox{270}{$B$}}"{above,pos=4.85}]{a}
        \end{quantikz}
    };
    \node[BrickRed] at (5.55,-1.23) {\meterDiag};
  \end{tikzpicture}
}

\newcommand{\firewallBob}{
  \begin{tikzpicture}
    \fill[color=BrickRed!30, dotted, thick, rounded corners] (3,-0.5) rectangle (9,2.5);
    \fill[color=tintedcolor, dotted, thick](7,1) circle (0.75);
    \draw[color=black, dotted, thick](7,1) circle (0.75);
    \draw[color=BrickRed, thick](7,1) circle (0.8);
    \node[BrickRed] at (5.8,0.75) {$\underbrace{R'}_{\quad   B \, \ \ H}$};
    \node[BrickRed] at (3.3,1) {$R$};
    \draw[thick, decorate, decoration={snake, amplitude=0.04cm, segment length=0.2cm}] 
    (5.6,1) -- (3.4,1);
    \node[] at (1.5,1) {\BobCircuit};
  \end{tikzpicture}
}

\makeindex             

\abstract{
The firewall paradox, a puzzle in black hole physics, depends on an implicit assumption: a rule that allows the infalling and the outside observer to combine their perspectives. However, a recent extension of the Wigner's friend paradox shows that such a combination rule conflicts with quantum theory --- without involving gravity. This challenges the usual conclusion of the firewall paradox, that standard quantum gravity assumptions are incompatible. More generally, black hole puzzles and Wigner's friend puzzles are closely related by a correspondence. This suggests that the firewall paradox may be a symptom of the same fundamental issue that leads to the extended Wigner's friend paradox. 
}

\begin{document}
\newgeometry{left=6.05cm, right=1.05cm, top=5cm, bottom=2cm}
\maketitle

\newpage
\restoregeometry
\section{Introduction}

\noindent Imagine two quantum physicists, Alice and Bob. Alice is enclosed in an isolated laboratory, where she performs a projective measurement on a qubit. From her perspective, after observing the outcome of her measurement, the qubit is in a pure state. Meanwhile, Bob, who stays outside, describes the entire laboratory --- including everything inside it --- as a big quantum system. From his perspective, this system undergoes a unitary time evolution, which generally entangles the qubit with other systems within the laboratory. Thus, the two physicists arrive at incompatible conclusions about the final state of the qubit: For Alice, it is in a pure state, whereas, for Bob, it is entangled with other systems. This is the basic version of Wigner's friend paradox~\cite{Wigner_1962}.\footnote{The physicist in Alice's role is commonly referred to as ``the Friend'' and the one in Bob's role as ``Wigner''. In quantum foundations, they are sometimes called ``agents'', a term associated with additional assumptions like ``free will'', which play no role in our arguments. In gravity, Alice and Bob are also called ``observers'', but this notion degrades them to a too passive role. Here, we want to emphasize the aspect that they can interact with physical systems and use a physical theory to make predictions. Therefore, we refer to them as ``physicists''.} 

Thought experiments in quantum gravity, like the Xeroxing paradox~\cite{Susskind_1994,Hayden_2007} and the firewall paradox~\cite{Mathur_2009,Almheiri_2013}, similarly involve different physicists adopting different perspectives. For example, Alice, who is freely falling into a black hole, has access to the interior, whereas Bob, who remains outside, has access to the Hawking radiation emitted after Alice reaches the horizon. Furthermore, as in Wigner's friend paradox, the conclusions that Alice and Bob draw about the state of certain systems, such as the radiation modes in ``the zone'', are incompatible. 

In this chapter, we aim to turn these superficial similarities between quantum foundations and quantum gravity considerations into a deeper correspondence. For this, we explore recent experiments that extend Wigner's friend paradox, developed within the field of quantum foundations~\cite{Deutsch_1985,Brukner_2017,Bong_2020,Frauchiger_2018} (for reviews see \cite{Nurgalieva_2020,Bub_2021,Schmid_2024}). They rely on the principle that ``physicists are physical''\footnote{The similarity to Landauer's slogan ``Information is physical''~\cite{Landauer_1991} is deliberate.}. This principle was already used in the basic version: it allowed Bob to apply quantum theory to Alice.  

Wigner's friend experiments, like quantum gravity thought experiments, test the universality of quantum theory --- the assumption that we can always use quantum theory to describe any physical system. Combined with the aforementioned principle, universality implies that quantum physicists can use quantum theory to describe other quantum physicists. As we will explain in \cref{sec:Wigner}, this test shows that 
quantum theory being universal is incompatible with seemingly innocent assumptions about how different physicists combine their conclusions. 

An example of such a rule is the following~\cite{Frauchiger_2018}.
Suppose that Bob, using quantum theory, concluded that another physicist, Charly, using quantum theory, concluded that the result of a measurement of a qubit is~$0$. The assumption is now that Bob can conclude that the measurement result is~$0$.

Combining information acquired by different physicists with different perspectives is also relevant to quantum gravity thought experiments, as we will discuss in \cref{sec:blackholes}. In situations where these pieces are not operationally accessible to a single physicist, as in the Xeroxing paradox, this issue is explicitly recognized as problematic and referred to as \emph{black hole complementarity}~\cite{Susskind_1993,Susskind_1994,Stephens_1994}. However, in other situations, particularly in the firewall paradox, the rules for combining information appeared so innocent that they remained implicit and unexamined. Making these rules explicit reveals that the firewall paradox belongs to same class of thought experiments as Wigner's friend paradox --- the class of thought experiments where the universality of quantum theory, together with a rule for combining information held by different physicists, leads to a contradiction.\footnote{Rules for combining information are also relevant in other theories, such as classical thermodynamics (see \cite{Jones_2025} for more examples). An example is Maxwell's demon paradox~\cite{Maxwell_1871}. It features a physicist, $\mathrm{P}$, who holds the usual coarse-grained thermodynamic description of a gas, and a demon, $\mathrm{D}$, who can observe individual gas molecules. 
A careless combination of these perspectives leads to the conclusion that $\mathrm{P}$ would observe a violation of the second law of thermodynamics \cite{Bennett_1982}. \label{footnote:Maxwell}}

\section{Wigner's friend thought experiments} \label{sec:Wigner}
For a long time, the discussion around Wigner's friend paradox was mostly philosophical, and the implications for concrete physics questions were unclear. 
Recent extensions of the experiment, proposed in \cite{Brukner_2017,Bong_2020,Frauchiger_2018}, changed this as they test specific properties of a physical theory, similarly to Bell experiments testing local causality \cite{Einstein_1935,Bell_1964}.
One of these properties is the universal applicability of the theory --- the tenet that any physicist can apply the theory to describe any systems around them. Wigner's friend thought experiments push the universality of quantum theory to its limits, by applying the theory to its users. 

Physicists are part of the world and, therefore, physical systems, which may again be described by other physicists. We frame this as a principle, which any fundamental physical theory, like quantum theory, must satisfy.

\begin{tcolorbox}[colback=tintedcolor,colframe=tintedcolor]
  \begin{customPrinciple}{(PP)}\label{principle:physical}
    --- \phantom{.}\emph{\textbf{Physicists are physical!}}

    \noindent
    A physicist can use the theory to describe another physicist who uses the theory.
  \end{customPrinciple}
\end{tcolorbox}

The principle treats physicists as objects of study. At first sight, this may appear problematic, potentially involving vague notions like ``free will'', ``consciousness'', or ``the mind''.\footnote{Indeed, Wigner's original conclusion was that a ``being with a consciousness must have a different role in quantum mechanics than the inanimate measuring device'' \cite{Wigner_1962}.}
However, we avoid this vagueness by adopting an operational approach and treating physicists as information-processing devices, for example, Turing machines connected to experimental devices.\footnote{\Cref{principle:physical} may be stated in a way reminiscent of the Church-Turing thesis \cite{Church_1936,Turing_1937}: A computer programmed with the rules of the theory can be modelled within the theory by another computer.}
The physicists' task is then captured by an algorithm for processing data and generating predictions using specific rules.\footnote{\Cref{principle:physical} also plays a central role in resolving Maxwell's demon paradox described in footnote~\ref{footnote:Maxwell}. On the one hand, to fulfil his task, the demon $\mathrm{D}$ needs to process information gained by observing the gas molecules. On the other hand, $\mathrm{D}$, and in particular his memory containing the record of his observations, must be modelled as a physical system. In particular, in $\mathrm{P}$'s coarse-grained description, $\mathrm{D}$ is itself subject to the laws of thermodynamics. Taking this into account, no violation of the second law occurs~\cite{Bennett_1982}.}

\subsection*{Quantum states cannot be agreed on\dots}
\begin{figure}
  \small
  \begin{tcolorbox}[colback=tintedcolor,colframe=tintedcolor]
    \begin{protocol}[Wigner's friend paradox \cite{Wigner_1962} with minor modifications]\label{ex:Wigner}
      \phantom{v}

      \noindent
      \textbf{Setup:}
      \begin{itemize}
        \item Bob is outside a laboratory. He has full quantum control over the laboratory as well as any system initially entangled with it. 
        He can send a qubit $Q$ into the laboratory and isolate the laboratory from its environment. 
        \item Alice is part of the laboratory and can control the qubit $Q$ she receives from the outside. We denote the part of the laboratory without the qubit by $L_A$.
      \end{itemize}
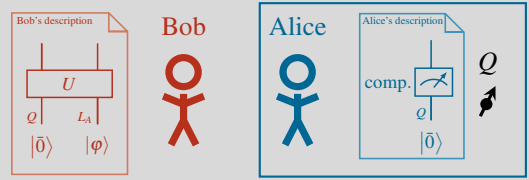
      \begin{center}
        \begin{tikzpicture}
          \node[text=BrickRed] (A) at (-3,1) {Bob}; 
          \node[text=BrickRed] (A) at (-3,0) {\Strichmaxerl[5] }; 
          \node[document,
          draw=BrickRed!60,
          semithick,,
          minimum width=15mm,
          minimum height=21.2mm,
          font=\ttfamily\Large, document dog ear=7pt] at (-4.5,0.1) {};
          \node[] at (-4.5,0) {\wignerCircuit};
          \node[text=BrickRed] at (-4.7,1.05) {\fontsize{4}{1}\selectfont{Bob's description}};
    
          \node[text=MidnightBlue] (A) at (-1.5,1) {Alice}; 
          \node[rotate=0,text=MidnightBlue] (A) at (-1.5,0) {\Strichmaxerl[5] }; 
          \draw[MidnightBlue, thick] (-2,-1) rectangle (1.5,1.3);
          \node (A) at (1,0.5) {$Q$}; 
          \node (A) at (1,0) {\qubit}; 
          \node[document,
          draw=MidnightBlue!60,
          semithick,
          minimum width=13.5mm,
          minimum height=19mm,
          font=\ttfamily\Large, document dog ear=7pt] at (-0,0.2) {};
          \node[] at (-0,0.05) {\friendCircuit};
          \node[text=MidnightBlue] at (-0.12,1.05) {\fontsize{4}{1}\selectfont{Alice's description}};
        \end{tikzpicture}
      \end{center}

      \noindent
      \textbf{Protocol:}
      \begin{enumerate}
        \item Bob prepares $L_A$ in a pure state $\ket{\varphi}_{L_A}$.\footnotemark 
        \item Bob prepares~$Q$ in state 
        \begin{equation}
          \ket{\bar{0}}_Q = \textstyle{\sqrt{\frac{1}{2}}}\big(\ket{0}_Q + \ket{1}_Q\big)
        \end{equation}
        and sends it to Alice into the laboratory.
        \item Bob isolates the laboratory.
        \item Alice measures~$Q$ projectively in the computational basis, $\{\ket{0}, \ket{1}\}$, and records the outcome.
        \item Alice infers the quantum state of $Q$.
        \item \label{item:Bob_inference} Bob infers the quantum state of the composite system $L_A Q$. 
      \end{enumerate}
    \end{protocol}
    \end{tcolorbox}
\end{figure}

We are now ready to examine the basic version of Wigner's friend paradox in more detail. 
The paradox involves two physicists, Alice and Bob, who follow the instructions specified in \cref{ex:Wigner}.

\footnotetext{
  If one assumes that the total state is pure, Bob can achieve this by an appropriate measurement of the systems initially entangled with the laboratory.
  Such a measurement does not act on the laboratory itself. 
  In particular, by non-signalling, it does not harm Alice. 
  The same is also true for the remaining steps. 
  For example, Bob can isolate the laboratory without acting on Alice. 
}

\Cref{principle:physical} plays a key role in this experiment: Alice is, at the same time, the subject using quantum theory and the object described by it.\footnote{That Alice is a physicist, able to use quantum theory, is the crucial distinction between Wigner's friend and Schrödinger's cat experiment. (Unlike the cat, she is also not subjected to a lethal dose of hydrocyanic acid~\cite{Schrodinger_1935}.)} This differs from standard experiments, where physicists are exterior to the experimental setup described by the theory. 

\Cref{ex:Wigner} requires both Alice and Bob to reason about the experiment, using the rules of quantum theory. However, their descriptions encompass different systems. Therefore, we take precautionary measures: We always declare explicitly from whose perspective a system is described. 
For example, we write $\smash{\rho^{\mathrm{Alice}}_Q}$ for the state Alice assigns to the qubit~$Q$. 
Additionally, we state the assumption that Alice and Bob are justified in their use of quantum theory explicitly.

\begin{customAssumption}{(Q)}\label{assumption:quantum}
  Any physicist can apply quantum theory\footnotemark \footnotetext{
    By quantum theory, we mean the postulates stated in \cite{Nielsen_2012}.
  } 
  to any physical system\footnotemark.\footnotetext{
    One might wonder whether this leads to problems if the system is not closed, such as a qubit~$A$ coupled to an environment $B$.
    Suppose that, in a particular run of an experiment, a physicist $\mathrm{P}_1$ has acquired information about $B$ that allows him to predict that $AB$ evolves into $\smash{\rho_{AB}^{\mathrm{P}_1}} = \ketbra{0}_A \otimes \ketbra{\varphi}_B$.
    Another physicist $\mathrm{P}_2$, who holds a description of qubit~$A$ only, may conclude that
    $A$ evolves to $\smash{\rho_A^{\mathrm{P}_2}} = \smash{\frac{\mathbbm{1}_A}{2}}$. 
    This mixed state reflects the fact that $\mathrm{P}_2$ has no knowledge about the environment $B$, and the coupling between $A$ and $B$ made this uncertainty propagate into $A$. 
    However, there is no contradiction between $\mathrm{P}_1$ and $\mathrm{P}_2$'s conclusions, as there exists a joint $\smash{\psi_{AB}} = \ketbra{0}_A \otimes \ketbra{\varphi}_B$ they could both agree on, according to the definition given below. 
  }
\end{customAssumption}

Before proceeding with the analysis of \cref{ex:Wigner}, we highlight two key aspects of a theory's universality. Each aspect corresponds to one of the words ``any'' in \cref{assumption:quantum}. First, any physicist can apply the theory, here, quantum theory. In particular, Alice, who 
is in an isolated laboratory, can apply quantum theory. Second, any system can be described by the theory. In particular, Bob can describe Alice as a quantum system. 

The first aspect of universality is often left implicit, although there are firm views on it.
Proponents of many-worlds~\cite{Vaidmann_2021} or Bohmian mechanics~\cite{Teufel_2009}, for instance, might proclaim that quantum theory can ultimately be applied only from a perspective outside the universe.\footnote{These approaches require that quantum theory be applied to a closed system. But when a physicist~$\mathrm{P}$ interacts with a system $S$, any closed system that includes $S$ will necessarily also include~$\mathrm{P}$. Consequently, $\mathrm{P}$ must refer to a hypothetical outside perspective onto this closed system. In many-worlds this perspective is captured by ``the wave function of the universe''. But this leads to an information-theoretic problem: $\mathrm{P}$ is fundamentally lacking the information to infer this wave function, because his access is limited to a single branch of it.} A similar view is common in AdS/CFT, where the boundary theory of the AdS universe is regarded as the ultimate authority. In contrast, \cref{assumption:quantum} asserts that quantum theory must be usable by the physicists inside the universe. This is a natural and hard-to-dispute requirement --- after all, the theory was discovered and validated through countless experiments by such physicists. 

We are now ready to analyse the descriptions of the qubit $Q$ that Alice and Bob arrive at by the end of \cref{ex:Wigner}. After having performed her projective measurement, and conditioned on the recorded outcome, Alice's state will be either $\smash{\ket{0}_Q}$ or $\smash{\ket{1}_Q}$, i.e., $\smash{\rho^{\mathrm{Alice}}_Q}$ is pure. Meanwhile, for Bob, who does not know Alice's measurement outcome, $Q$ will be in a mixed state, i.e., $\smash{\rho^{\mathrm{Bob}}_Q} = \smash{\frac{1}{2}\mathbbm{1}_Q}$. Furthermore, Bob can apply quantum theory to the entire laboratory $L_A Q$, which, for him, is in a pure state. Because it is isolated and thus evolves unitarily, it will remain pure. But since the marginal state $\smash{\rho^{\mathrm{Bob}}_Q}$ is mixed, he concludes that~$Q$ is entangled with the rest of the laboratory $L_A$. 

One may now ask whether Alice's and Bob's conclusions can be combined, in the sense that there exists a quantum state, $\psi$, on which both of them can agree. We say that a physicist $\mathrm{P}$ \emph{can agree on $\psi$} if, for any finite-dimensional system~$S$ described by $\mathrm{P}$, there exists information such that, conditioned on this information, the state $\smash{\rho_S^{\mathrm{P}}}$ that~$\mathrm{P}$ assigns to $S$ equals $\psi_S$. It is not hard to convince oneself that this corresponds to the condition $\smash{\mathrm{supp}(\psi_S) \subseteq \mathrm{supp}(\rho_S^{\mathrm{P}})}$, where $\mathrm{supp}(\sigma)$ denotes the support of the operator $\sigma$. Note that, if $\smash{\rho_S^{\mathrm{P}}}$ is pure then this condition implies $\smash{\psi_S} = \smash{\rho_S^{\mathrm{P}}}$. 

Let us apply this definition to Alice's and Bob's descriptions. Our analysis above showed that for Alice the qubit $Q$ is in a pure state $\smash{\rho_Q^{\mathrm{Alice}}}$, whereas for Bob the joint system $QL_A$ is in an entangled state $\smash{\rho_{QL_A}^{\mathrm{Bob}}}$, which is also pure. As we just noted, the state~$\psi$ would need to satisfy both $\psi_Q = \smash{\rho_Q^{\mathrm{Alice}}}$ and $\psi_{QL_A} = \smash{\rho_{QL_A}^{\mathrm{Bob}}}$. This is impossible because, according to $\psi$, the qubit $Q$ would at the same time be pure and entangled with $L_A$. Therefore, there is no state $\psi$ on which both Alice and Bob can agree. We formulate this conclusion as a no-go theorem: 

\begin{theorem}[State agreement paradox] \label{thm:obj_state}
  These assumptions are incompatible: 
  \begin{itemize}
    \item {\textbf{Executability:}} Physicists can execute \cref{ex:Wigner}.
    \item {\textbf{Universal applicability of quantum theory:}} \Cref{assumption:quantum} holds.
    \item {\textbf{State agreement:}} For any two physicists $\mathrm{P}$ and $\mathrm{P'}$ who use quantum theory there exists a state $\smash{\psi}$ on which both $\mathrm{P}$ and $\mathrm{P'}$ can agree.
  \end{itemize}
\end{theorem}

If Alice and Bob want to agree on the quantum state of $Q$, then at least one of them has to admit a mistake. We now discuss the two possible options.\footnote{Combinatorially, there are three, but when Alice and Bob are both wrong, this is akin to admitting that quantum theory is not applicable from the perspective of any physicist.} 

If Alice is correct and Bob is wrong, this indicates that physics imposes a constraint on at least one of the first two assumptions, preventing Bob from describing and controlling large systems such as Alice's laboratory. Justifying such a constraint is the program pursued by objective collapse theorists \cite{Ghirardi_1986}. They propose to modify quantum theory and supplement it with a collapse mechanism. The argument leading to \cref{thm:obj_state} can be used to impose bounds on the strength of this mechanism: it must be strong enough to prevent the conclusions Bob draws about Alice, who may be a quantum computer programmed with the rules of quantum theory.\footnote{Although the algorithm that implements the rules is classical and therefore can run on a classical computer, Bob needs to have quantum control over the computer.} Conversely, the collapse mechanism must be weak enough to not be ruled out by current experiments. 
This leaves only a small window of possible collapse models, which, in view of recent progress in quantum technologies, is likely to close completely.

If Alice is wrong and Bob is correct, then this indicates a mechanism that prevents physicists in isolated laboratories, like Alice, from applying quantum theory. This option entails another modification of \cref{assumption:quantum}. For example, one could demand that only physicists who are not in superposition can use quantum theory. However, this is problematic, as we cannot determine whether we are in such a situation:
Consider a powerful alien physicist who would describe our entire galaxy and any systems entangled with it. From the alien's perspective, we, together with Alice and Bob, could be in a superposition of performing the experiment or not performing it. Hence, the modified rule would not allow us to use quantum theory.
Nonetheless, this is the approach taken by many-worlds and Bohmian mechanics. They indeed consider an ultimate outside perspective.\footnote{While reasoning about such a perspective might be of interest for theological reasons, it stands in stark contrast to the fact that quantum theory was discovered and experimentally confirmed by us who inhabit the universe.} 

The first two assumptions can be experimentally tested and are not unlikely to be confirmed.\footnote{Such experimental tests still require some assumptions, e.g., that if another physicist falsifies \cref{assumption:quantum}, we should also consider this a falsification of \cref{assumption:quantum}.} This would mean we have to abolish the third assumption of \cref{thm:obj_state}. Then, both Alice and Bob can be correct even though they do not agree in their state assignment. Consequently, their states cannot represent an objective property of the qubit $Q$. But if states are not objective, is there anything that we can firmly say about the world? 

We may hope that, even if states are not objective, it is still possible to agree on things that can be directly observed --- measurement outcomes. We will explore this possibility with more elaborate Wigner's friend experiments. But before doing so, we highlight two information-theoretic aspects of quantum theory. 

The first is the \emph{state update rule}. The state assignment of a physicist $\mathrm{P}$ changes as she obtains new information about a system, such as the outcome of a measurement. Let $\smash{\rho^{\mathrm{P}}_{QS}}$ be the state that $\mathrm{P}$ assigns to a joint system $QS$. If $\mathrm{P}$ measures the subsystem $Q$ and obtains outcome $x$, then the \emph{updated state} that $\mathrm{P}$ assigns to $S$ is
\begin{equation}\label{eq:update_rule}
  \sigma^{\mathrm{P}|Q = x}_{S} \propto \tr_Q(\ketbra{x}_Q \rho_{QS}^{\mathrm{P}})
\end{equation}
where $\ketbra{x}_Q$ is the projector corresponding to this measurement outcome. Note that we extended the superscript to keep track of the outcome, which is now stored in $\mathrm{P}$'s memory. 

The second concept is the \emph{Heisenberg cut} of a physicist \cite{Heisenberg_1985}. 
It defines the boundary separating the systems that a physicist models within the theory from those she does not. Alice's Heisenberg cut, for instance, surrounds the qubit $Q$, whereas Bob's cut surrounds the entire laboratory. 

In a universal theory like quantum theory, the location of the Heisenberg cut is not determined by the theory. Indeed, \cref{assumption:quantum} allows each physicist to choose her cut arbitrarily. However, to avoid problems of self-reference, we will not consider scenarios where a physicist is enclosed within her own Heisenberg cut, as this would require her to describe herself. 

\subsection*{\dots{}outcomes are not objective\dots}

Deutsch proposed an extension of Wigner's experiment~\cite{Deutsch_1985}, \cref{ex:Deutsch}, in which a qubit is measured in two complementary bases. This, by itself, is not in conflict with quantum theory, as the measurement to a single physicist. However, we can use this experiment to prove a stronger no-go result, \cref{thm:obj_meas}, which questions the objectivity of measurement outcomes.

\begin{figure}
\small
\begin{tcolorbox}[colback=tintedcolor,colframe=tintedcolor]
  \begin{protocol}[Deutsch's extension of Wigner's friend paradox \cite{Deutsch_1985}]\label{ex:Deutsch}
    \phantom{v}

    \noindent
    \textbf{Setup:}
    \begin{itemize}
      \item Bob is outside a laboratory, over which he has full quantum control. He can send a qubit
      $Q$ into the laboratory and isolate the laboratory from its environment. 
      \item Alice is part of the laboratory and can control the qubit~$Q$ she receives from the outside. We denote the part of the laboratory without the qubit by $L_A$.
    \end{itemize}

    \begin{center}
      \begin{tikzpicture}
          \node[text=BrickRed] (A) at (-3,1) {Bob}; 
          \node[text=BrickRed] (A) at (-3,0) {\Strichmaxerl[5] }; 
          \node[] at (-5.1,0) {\bobDeutschCircuit};
          \node[document,
          draw=BrickRed!60,
          semithick,
          minimum width=25mm,
          minimum height=35.4mm,
          font=\ttfamily\Large, document dog ear=7pt] at (-5,0.1) {};
          \node[text=BrickRed] at (-5.73,1.77) {\fontsize{4}{1}\selectfont{Bob's description}};

          \node[text=MidnightBlue] (A) at (-1.5,1) {Alice}; 
          \node[rotate=0,text=MidnightBlue] (A) at (-1.5,0) {\Strichmaxerl[5] }; 
          \draw[MidnightBlue, thick] (-2,-1) rectangle (1.5,1.3);
          \node (A) at (1,0.5) {$Q$}; 
          \node (A) at (1,0) {\qubit}; 
          \node[document,
          draw=MidnightBlue!60,
          semithick,
          minimum width=13.5mm,
          minimum height=19mm,
          font=\ttfamily\Large, document dog ear=7pt] at (-0,0.2) {};
          \node[] at (-0,0.05) {\friendDeutschCircuit};
          \node[text=MidnightBlue] at (-0.12,1.05) {\fontsize{4}{1}\selectfont{Alice's description}};
      \end{tikzpicture}
  \end{center}

    \noindent
    \textbf{Protocol:}
    \begin{enumerate}
      \item Bob prepares~$Q$ and sends it to Alice into the laboratory. 
      \item\label{item:Bob_isolation} Bob isolates the laboratory.
      \item Alice measures~$Q$ in the computational basis and records the outcome in register $R_A$.
      \item Bob reverses the unitary evolution of $L_A Q$ back to step~\ref{item:Bob_isolation}.\textsuperscript{\ref{fn:aliceSurvival}}
    
      \item Bob measures~$Q$ in the diagonal basis, defined by 
      \begin{equation}
        \ket{\bar{0}}_Q =  \textstyle{\sqrt{\frac{1}{2}}}\left(\ket{0}_Q + \ket{1}_Q\right) \text{ and } \ket{\bar{1}}_Q = \textstyle{\sqrt{\frac{1}{2}}}\left(\ket{0}_Q - \ket{1}_Q\right),
      \end{equation}
      and records the outcome in $R_B$. 
    \end{enumerate}
  \end{protocol}
\end{tcolorbox}
\end{figure}

\refstepcounter{footnote}\label{fn:aliceSurvival} \footnotetext{
  This is a highly invasive operation on Alice. One may be worried that the argument presented here depend on Alice surviving this operation \cite{Aaronson_2018}. However, the argument does not rely on Alice being a physicist after this step.
}

To this end, we first need to define what we mean by \emph{objective measurement outcomes}. The idea is to use the following operational criterion: if measurement outcomes are objective, then there should be a rule which any physicist can use to consistently update her description of the experiment. Specifically, whenever a physicist's description is initially compatible with quantum theory, updating it with the objective outcomes must preserve this compatibility. 

Returning to \cref{ex:Deutsch}, let us assume by contradiction that there exists a rule for updating a description such that it incorporates both the outcome observed by Alice, $r_A$, and the outcome observed by Bob, $\bar{r}_B$. This rule may be represented as a function that takes as input any possible initial descriptions in the form of a quantum state $\rho_{QS}$ of the qubit~$Q$ and a reference system called $\smash{S}$, as well as the values of the outcomes $r_A$ and $\bar{r}_B$, 
\begin{align}
  \mathcal{R}: (\rho_{Q S}, r_A,\bar{r}_B) \mapsto (P^{\mathcal{R}}_{R_A,R_B}(r_A, \bar{r}_B), \sigma^{\mathcal{R}}_{S| R_A = r_A, R_B = \bar{r}_B}).
\end{align}
The outcome consists of two pieces. The first, $P^{\mathcal{R}}_{R_A,R_B}(r_A, \bar{r}_B)$, is the probability that the outcome pair $(r_A, \bar{r}_B)$ occurs, which is needed to allow updating if one has access to only one of the outcomes $r_A$ or $\bar{r}_B$. The second, $\smash{\sigma^{\mathcal{R}}_{S| R_A = r_A, \, R_B = \bar{r}_B}}$, is the updated quantum description. The only requirement we impose on the update rule $\mathcal{R}$ is that, it is compatible with the quantum-theoretic update rule~\eqref{eq:update_rule}. For example, because Alice has access to $r_A$, we require that 
\begin{align}\label{eq:req_Alice}
  \forall r_A: \quad
  \sum_{\bar{r}_B} P^{\mathcal{R}}_{R_A,R_B}(r_A,\bar{r}_B) \, \sigma^{\mathcal{R}}_{S| R_A = r_A, R_B = \bar{r}_B}  &= \tr_Q(\ketbra{r_A}{r_A}_Q \rho_{QS}).
\end{align}
Here, $\ketbra{r_A}{r_A}_Q$ is the projector onto a computational basis state, corresponding to Alice's measurement. 

To show that such an update rule $\mathcal{R}$ cannot exist, we rewrite it as the function
\begin{align}
  \mathcal{F}: \rho_{Q S} \mapsto \sigma_{R_A R_B S} = \sum_{r_A,\bar{r}_B} P^{\mathcal{R}}_{R_A R_B}(r_A,\bar{r}_B) \ketbra{r_A}{r_A}_{R_A} \otimes \ketbra{\bar{r}_B}{\bar{r}_B}_{R_B} \otimes \sigma^{\mathcal{R}}_{S| R_A = r_A, \, R_B = \bar{r}_B}.
\end{align}
Note that, while $\mathcal{F}$ is not necessarily a linear map, its output is by construction a valid quantum state. We can, therefore, rewrite both the compatibility condition for updating on $r_A$, \cref{eq:req_Alice}, and the compatibility condition for updating on $r_B$ as
\begin{align}\label{eq:nice_condition}
  \begin{split}
    \tikz[baseline=0ex,decoration={
      markings,
      mark=at position 0 with {\arrow{latex}}}]{
      \draw[thick] (-1.25,0.25) -- (-0.5,0.25);
      \draw[thick] (-1.25,-0.25) -- (-0.5,-0.25);
      \node (A) at (-0.6,0.4) {\scriptsize{$Q$}}; 
      \node (A) at (-0.6,-0.1) {\scriptsize{$S$}}; 
      }\xmapsto{\tr_{R_B} \circ \mathcal{F}}
      \tikz[baseline=0ex]{
        \draw[thick] (1.5,-0.25) -- (2.25,-0.25);
        \draw[double] (1.5,0.25) -- (2.25,0.25);
      \node (A) at (1.65,0.4) {\scriptsize{$R_A$}}; 
      \node (A) at (1.65,-0.1) {\scriptsize{$S$}};
    }  
    &=   
    \tikz[baseline=0ex]{
      \draw[double] (1,0.25) -- (1.5,0.25);
      \draw[thick] (-0.5,0.25) -- (-1,0.25);
      \draw[thick] (-0.5,-0) rectangle (1,0.5);
      \node (A) at (0.25,0.25) {$\mathcal{Z}_{R_A|Q}$}; 
      \node (A) at (-0.95,0.4) {\scriptsize{$Q$}}; 
      \node (A) at (1.4,0.4) {\scriptsize{$R_A$}}; 
      \node (A) at (-0.95,-0.1) {\scriptsize{$S$}}; 
      \draw[thick] (-1,-0.25) -- (1.5,-0.25);
    } \\ \\
    \tikz[baseline=0ex,decoration={
      markings,
      mark=at position 0 with {\arrow{latex}}}]{
      \draw[thick] (-1.25,0.25) -- (-0.5,0.25);
      \draw[thick] (-1.25,-0.25) -- (-0.5,-0.25);
      \node (A) at (-0.6,0.4) {\scriptsize{$Q$}}; 
      \node (A) at (-0.6,-0.1) {\scriptsize{$S$}}; 
      }\xmapsto{\tr_{R_A} \circ \mathcal{F}}
      \tikz[baseline=0ex]{
      \draw[thick] (1.5,-0.25) -- (2.25,-0.25);
      \draw[double] (1.5,0.25) -- (2.25,0.25);
      \node (A) at (1.65,0.4) {\scriptsize{$R_B$}}; 
      \node (A) at (1.65,-0.1) {\scriptsize{$S$}};
    } 
    &=   
    \tikz[baseline=0ex]{
      \draw[double] (1,0.25) -- (1.5,0.25);
      \draw[thick] (-0.5,0.25) -- (-1,0.25);
      \draw[thick] (-0.5,-0) rectangle (1,0.5);
      \node (A) at (0.25,0.25) {$\mathcal{X}_{R_B|Q}$}; 
      \node (A) at (-0.95,0.4) {\scriptsize{$Q$}}; 
      \node (A) at (1.4,0.4) {\scriptsize{$R_B$}}; 
      \node (A) at (-0.95,-0.1) {\scriptsize{$S$}}; 
      \draw[thick] (-1,-0.25) -- (1.5,-0.25);
    }
  \end{split}
\end{align}
where $\mathcal{Z}_{R_A|Q}$ and $\mathcal{X}_{R_B|Q}$ are completely positive trace-preserving maps corresponding to the measurements in the computational and diagonal basis, respectively.

But the two conditions in \cref{eq:nice_condition} cannot hold simultaneously. To see this, we may choose the reference system $S$ to be a qubit maximally entangled with $Q$, i.e., $\smash{\rho_{Q S}} = \smash{\ketbra{\psi}{\psi}_{Q S}}$ with $\ket{\psi}_{QS} \propto \smash{\ket{0}_Q \ket{0}_S + \ket{1}_Q \ket{1}_S}$. In this case, $\mathcal{F}$ produces the output $\smash{\sigma_{R_A R_B S}} \coloneq \mathcal{F}(\rho_{Q S})$. We now calculate the conditional entropies of the outcomes $Z$ and $X$ of a measurement of $S$ in the computational basis and in the diagonal basis, conditioned on $R_A$ and $R_B$, respectively. Because these entropies only depend on the marginals $\smash{\sigma_{R_B S}}$ and $\smash{\sigma_{R_A S}}$, we can use \cref{eq:nice_condition} to see that 
\begin{equation} \label{eq:complementary_entropies}
    H(Z|R_A) = H(X|R_B) = 0.
\end{equation}
But this contradicts strong subadditivity and the entropic Heisenberg uncertainty principle~\cite{Maassen_1988}, which assert that 
\begin{equation} \label{eq:Heisenberg_uncertainty}
  H(Z|R_A) + H(X|R_B) \geq H(Z|R_A R_B) + H(X|R_A R_B)  \geq 1.
\end{equation}
Therefore, our assumption that an update function $\mathcal{R}$ exists must be wrong.\footnote{This violation --- expressed by the discrepancy between \cref{eq:complementary_entropies} and \cref{eq:Heisenberg_uncertainty} --- is robust, meaning that even an approximate version of $\mathcal{R}$ cannot exist.} 

We summarize this conclusion as another no-go theorem.\footnote{Similar no-go theorems have been proposed in \cite{Brukner_2017,Bong_2020}. These works consider a different experiment which, in contrast to ours, relies on the idea that measurement settings can be chosen freely. Their notion of objectivity is then defined based on free choice: they demand that, conditioned on any choice of the measurement setting, there is a joint probability distribution of all measurement outcomes that satisfies certain assumptions. Specifically, these are no-superdeterminism and locality, analogously to the assumptions that enter Bell's theorem.}

\begin{theorem}[Objective outcome paradox]\label{thm:obj_meas}
  These assumptions are incompatible:
  \begin{itemize}
    \item {\textbf{Executability:}} Physicists can execute \cref{ex:Deutsch}.
    \item {\textbf{Universal applicability of quantum theory:}} \Cref{assumption:quantum} holds. 
    \item {\textbf{Objective measurement outcomes:}} 
    A physicist's description of a physical system can be updated with all measurement outcomes ever observed.
  \end{itemize}
\end{theorem}

The purpose of our no-go theorems is to show that certain assumptions about physics, which \emph{a priori} sound reasonable, cannot hold.  The first two assumptions of \cref{thm:obj_meas} capture the universality of quantum theory, whereas the third can be regarded as a rule to combine information held by different physicists. We have already described earlier, after \cref{thm:obj_state}, that the validity of the first two assumptions is, in principle, experimentally testable. Provided that this test will confirm them,  \cref{thm:obj_meas} implies that the third assumption is wrong. 

Giving up the third assumption --- the objectivity of measurement outcomes --- might seem like a radical move. However, the assumption refers to outcomes that are not in general accessible to a single physicist, so from a purely operational point of view, there is no problem. \Cref{ex:Deutsch} illustrates this inaccessibility. If Bob asked Alice to tell him~$r_A$, he would break the laboratory's isolation, making its evolution non-unitary, and thus change his other conclusions. Conversely, the reversal of the time evolution of the laboratory will erase Alice's memory of $r_A$. Hence, even if Bob told her the outcome $\bar{r}_B$ after his final measurement, she no longer knows~$r_A$.\footnote{This can be seen as a form of complementarity, which we will discuss further in \cref{sec:blackholes}.} Consequently, the assumption of objective measurement outcomes has no operational relevance --- any argument defending it would be purely philosophical.

\subsection*{\dots{} and even communication cannot help}
Expecting that all measurement outcomes --- even those which are not operationally accessible --- can be consistently incorporated in any physicist's description might have still been too much to ask. Here, we derive another no-go result, the \emph{quantum collaboration paradox}, which we will state as \cref{thm:FR}. It replaces the assumption that measurement outcomes are objective by a weaker and more operational consistency assumption, phrased as \cref{assumption:consistency} below. To emphasize the operational nature of the argument, we present it by referring to a game, the \emph{complementarity game}.

\begin{namedgame}[Complementarity]\namedlabel{game:complementarity}{complementarity game}
  Consider $N$ collaborating physicists and a referee. The game is played as follows. 
  \begin{enumerate}
    \item The physicists indicate a qubit $Q$ to the referee.
    \item The physicists issue a pair of predictions $(P, \bar{P})$ to the referee. 
    \item Depending on a fair coin, the referee chooses one of two possible tests:
    \begin{enumerate}[topsep=0em]
      \item Measure $Q$ in the computational basis and check if $P$ equals the result.
      \item Measure $Q$ in the diagonal basis and check if $\bar{P}$ equals the result.
    \end{enumerate}
  \end{enumerate}
  The $N$ collaborating physicists win if the test of the referee succeeds. 
\end{namedgame}

\begin{figure}[!htb]
  \small
  \begin{tcolorbox}[colback=tintedcolor,colframe=tintedcolor]
  \begin{protocol}[Wigner's friend strategy]\label{ex:consistency}
  \phantom{v}

  \noindent
    \textbf{Setup:}
    \begin{itemize}
      \item Darwin is outside two laboratories, $L_A$ and $L_C$. He has quantum control over $L_C$
      and can send a qubit $Q_A$ into $L_A$ and a qubit $Q_C$ into $L_C$. 
      \item Bob is outside the laboratories $L_A$ and $L_C$ and has quantum control over $L_A$.
      \item Alice is part of laboratory $L_A$ and can control the qubit $Q_A$. 
      \item Charly is part of laboratory $L_C$ and can control the qubit $Q_C$. 
    \end{itemize}
    
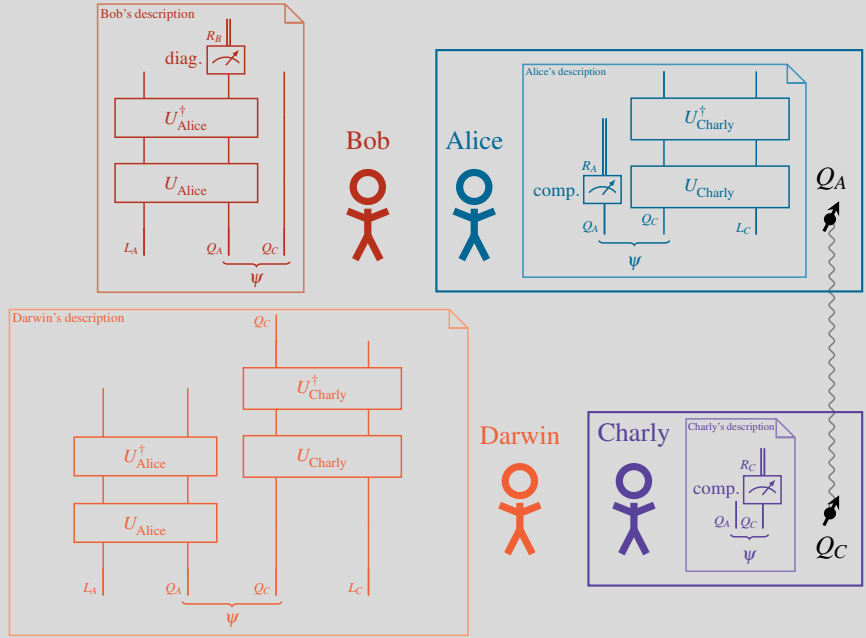
    \begin{center}
    \begin{tikzpicture}
      \draw[semithick, decorate, decoration={snake, amplitude=0.035cm, segment length=0.19cm}, gray] (1.1,-3.9)--(1.1,0);
    
      \node[text=MidnightBlue] (A) at (-3.6,1) {Alice}; 
      \node[rotate=0,text=MidnightBlue] (A) at (-3.6,0) {\Strichmaxerl[5] }; 
      \draw[MidnightBlue, thick] (-4.1,-1) rectangle (1.5,2.2);
      \node (A) at (1.1,0.5) {$Q_A$}; 
      \node (A) at (1.1,0) {\qubit}; 
      \node[document,
      draw=MidnightBlue!60,
      semithick,
      minimum width=37mm,
      minimum height=28mm,
      font=\ttfamily\Large, document dog ear=7pt] at (-1.1,0.6) {};
      \node[] at (-1.1,0.6) {\AliceCircuitFR};
      \node[text=MidnightBlue] at (-2.4,1.9) {\fontsize{4}{1}\selectfont{Alice's description}};
 
      \node[text=Violet] (A) at (-1.5,-2.9) {Charly}; 
      \node[rotate=0,text=Violet] (A) at (-1.5,-3.9) {\Strichmaxerl[5] }; 
      \draw[Violet, thick] (-2.1,-4.9) rectangle (1.5,-2.6);
      \node (A) at (1.1,-4.4) {$Q_C$}; 
      \node (A) at (1.1,-3.9) {\qubit}; 
      \node[document,
      draw=Violet!60,
      semithick,
      minimum width=14.1mm,
      minimum height=20mm,
      font=\ttfamily\Large, document dog ear=7pt] at (-0.1,-3.7) {};
      \node[] at (-0.1,-3.85) {\CharlyCircuitFR};
      \node[text=Violet] at (-0.22,-2.8) {\fontsize{4}{1}\selectfont{Charly's description}};

      \node[document,
      draw=BrickRed!60,
      semithick,
      minimum width=26.9mm,
      minimum height=38mm,
      font=\ttfamily\Large, document dog ear=7pt] at (-7.2,0.9) {};
      \node[text=BrickRed] (A) at (-5,1) {Bob}; 
      \node[text=BrickRed] (A) at (-5,0) {\Strichmaxerl[5] }; 
      \node[] at (-7,0.8) {\BobCircuitFR};
      \node[text=BrickRed] at (-7.9,2.67) {\fontsize{5}{1}\selectfont{Bob's description}};

      \node[document,
      draw=RedOrange!60,
      semithick,
      minimum width=60.1mm,
      minimum height=43mm,
      font=\ttfamily\Large, document dog ear=7pt] at (-6.7,-3.4) {};
      \node[text=RedOrange] (A) at (-3,-2.9) {Darwin}; 
      \node[text=RedOrange] (A) at (-3,-3.9) {\Strichmaxerl[5] }; 
      \node[] at (-6.5, -3.4) {\RefCircuitFR};
      \node[text=RedOrange] at (-8.94,-1.37) {\fontsize{5}{1}\selectfont{Darwin's description}};
    \end{tikzpicture}
    \end{center}    
    \vspace{-0.3cm}
    \noindent
    \textbf{Protocol:}
    \begin{enumerate}
      \item\label{item:start} Darwin prepares $Q_A$ and $Q_C$ in state\footnotemark 
      \begin{equation} \label{eq:Hardystate}
        \ket{\psi}^{\mathrm{all}}_{Q_A Q_C} = \textstyle{\sqrt{\frac{1}{3}}}\left(\ket{0}_{Q_A}\ket{0}_{Q_C} + \ket{0}_{Q_A}\ket{1}_{Q_C} + \ket{1}_{Q_A}\ket{0}_{Q_C}\right)
      \end{equation}
      and sends them to Alice and Charly, respectively. 
      \item \label{item:Alice_Measurement} Alice measures $Q_A$ in the computational basis and records the outcome in register~$R_A$.
      \item \label{item:Charly_measurement} Charly measures $Q_C$ in the computational basis and records the outcome in $R_C$.
      \item \label{item:Bob_reversal} Bob reverses the evolution of $L_A Q_A$ back to the beginning of step~\ref{item:Alice_Measurement}. 
      \item \label{item:Bob_measurement} Bob measures $Q_A$ in the diagonal basis,
      records the outcome in $R_B$, and communicates it to Darwin.
      \item Bob continues if the outcome is $\bar{1}$, else he restarts the experiment from step~\ref{item:start}.
      \item Darwin reverses the evolution of $L_C Q_C$ back to the beginning of step~\ref{item:Charly_measurement}.
      \item\label{item:indication} Bob indicates the qubit $Q_C$ to the referee.
      \item \label{item:Bob_abort} Bob issues the pair of predictions $(P = 1,\bar{P} = \bar{0})$ to the referee.
    \end{enumerate}
  \end{protocol}
\end{tcolorbox}
\end{figure}
\footnotetext{This state is common knowledge to all physicists because we assume that all of them
are provided with a description of the protocol.}

In \cref{ex:consistency}, we describe a strategy, which is based on the universality of quantum theory and a consistency assumption, that allows the players to win the \ref{game:complementarity} with probability~$1$. This will lead to a contradiction, as according to quantum theory the maximal probability of winning this game is $\smash{\frac{1}{2} + \textstyle{\frac{1}{\sqrt{8}}}} < 1$ \cite[Eq.~(17) with $m = 2$]{Renes_2017}.

We analyse \cref{ex:consistency} from Bob's perspective. Bob will also reason about the viewpoints of other physicists, like Charly. He does this by simulating their reasoning using the information he has about them. 

\smallskip
\noindent
\textbf{Bob's reasoning:}
Bob chooses his Heisenberg cut to surround the joint system $L_A Q_A Q_C$ but not himself. Note that this is a valid choice, as it surrounds the laboratory $L_A$ and the qubit $Q_A$, which Bob must act on to reverse its time evolution. Furthermore, for Bob, the physicists Charly and Darwin are classical.

After step~\ref{item:Bob_reversal}, the state he assigns to the subsystem $Q_A R_C$ is 
\begin{equation}
  \rho^{\mathrm{Bob}}_{Q_A R_C} = \frac{1}{3}\left(2 \ketbra{\bar{0}}{\bar{0}}_{Q_A} \otimes \ketbra{0}{0}_{R_C} + \ketbra{0}{0}_{Q_A} \otimes \ketbra{1}{1}_{R_C}\right)
\end{equation}
Once Bob has indicated a qubit to the referee, i.e., he has reached step~\ref{item:indication}, his measurement result is $\bar{1}$.
The update rule~\eqref{eq:update_rule} yields
\begin{equation} \label{eq:update_conditioned_on_1}
  \rho^{\mathrm{Bob}|Q_A = \bar{1}}_{Q_C} = 6 \tr_{Q_A}\left(\ketbra{\bar{1}}{\bar{1}} \rho^{\mathrm{Bob}}_{Q_A R_C}\right)
  = \ketbra{1}{1}_{Q_C}.
\end{equation}
Bob is thus certain that Charly's measurement outcome was $1$. Note that, from \cref{eq:update_conditioned_on_1}, it also follows that the probability of Bob getting outcome $\bar{1}$ is strictly larger than $0$. This confirms that the protocol will indeed advance to step~\ref{item:indication} after finite time. Next Bob simulates Charly's reasoning using the information that Charly observed outcome~$1$.

\setlength{\leftskip}{1cm}
\noindent
\textbf{Bob's simulation of Charly's reasoning:}
Charly chooses his Heisenberg cut to surround $Q_A Q_C$ and such that all physicists are classical. Having observed the measurement outcome $1$, she can apply the update rule~\eqref{eq:update_rule} to the state~\eqref{eq:Hardystate}, making her assign the state $\smash{\ket{0}_{Q_A}}$ to $Q_A$. She is thus certain that Alice's outcome is $0$. 
To proceed, Charly simulates Alice's reasoning using this information.

\setlength{\leftskip}{2cm}
\noindent
\textbf{Charly's simulation of Alice's reasoning:}
Alice applies quantum theory to $Q_A L_C Q_C$\footnote{The system $L_C$ needs to be surrounded by Alice's Heisenberg cut, as she makes a statement about $Q_C$ after the referee has reversed Charly's measurement. One could, rightfully, be worried that Charly cannot do this simulation as she needs to simulate herself. One way to avoid this problem is for Alice to run the simulation of what she would do for all her possible measurement outcomes and communicate the result to Charly.}. Having observed the measurement outcome $0$, Alice can apply the update rule~\eqref{eq:update_rule} to \eqref{eq:Hardystate}, which leads her to assign the state $\smash{\ket{\bar{0}}_{Q_C}}$ to $Q_C$. She is thus certain that a measurement of the referee in the diagonal basis will yield outcome~$\bar{0}$.

\smallskip
\setlength{\leftskip}{0pt}

Based on this analysis, Bob has information about the information his fellow physicists possess. We summarize it here.\footnote{The derivation of the statements below requires Bob to combine the statements derived above using standard logic. There are researchers who think logic needs to be modified when applied in the context of quantum experiments \cite{Samuel_2023,Atzori_2024}. However, the language in which papers are written, even those who deny logic, use some form of logic. Hence, \cref{principle:physical} implies that physicists who argue that quantum theory breaks logic cannot themselves use logic in their arguments.}
\begin{enumerate}
  \item\label{item:Bob_Charly} Bob is certain that Charly's measurement has outcome~$1$ 
  \item\label{item:Bob_Charly_Alice} Bob is certain that Charly is certain that Alice's measurement has outcome~$0$.
  \item\label{item:Bob_Charly_Alice_ref} Bob is certain that Charly is certain that Alice is certain that the referee's measurement, if the diagonal basis was chosen, has outcome~$\bar{0}$.\footnote{Those worried about the counterfactual nature of this statement may consider a similar game where the referee's tests are deterministic.  Here, the goal is to predict a pair of values $(c, d)$ derived from the outcome $m$ of the referee's measurement by taking the logical AND with the bit that determines whether the measurement is in the computational $(b=0)$ or the diagonal ($b=1$) basis. Specifically, $(c, d)  = (\lnot b \text{ AND } \lnot m, b \text{ AND } m)$. Alice then predicts $d$, and the statement reads: Bob is certain that Charly is certain that Alice is certain that $d$ is~$0$.}
\end{enumerate}
These statements can be experimentally tested. For example, Bob could, in principle, abort the game before the Darwin undoes Charly's measurement, break the isolation of Charly's laboratory, and ask her what she knows about Alice's measurement. If quantum theory is correct, Bob would find statement~\ref{item:Bob_Charly_Alice} of the summary confirmed by this test. 

It seems Bob can use statement~\ref{item:Bob_Charly_Alice_ref} to be certain that with prediction $\bar{P} = \bar{0}$ he will win the game if the referee measures in the diagonal basis. 
However, to arrive at this conclusion, he needs an additional assumption:
\begin{customAssumption}{(C)}\label{assumption:consistency} 
  Let $\mathrm{P}_1$ be a physicist who describes another physicist $\mathrm{P}_2$ as a classical information-processing system.\footnote{We say that $\mathrm{P}_1$ describes a system as \emph{classical} if there exists an orthonormal basis with respect to which all states that $\mathrm{P}_1$ assigns to the system are diagonal. For a typical system that strongly interacts with its environment, this condition is met if the Heisenberg cut is placed between system and environment --- the system has decohered.} \\
  If $\mathrm{P}_1$ makes the statement ``I am certain that $\mathrm{P}_2$ is certain, based on reasoning using the same theory as me, that measurement $M$ has outcome~$x$'', then $\mathrm{P}_1$ can also make the statement ``I am certain that measurement $M$ has outcome~$x$''.
\end{customAssumption}
Using \cref{assumption:consistency}, Bob can now make the desired conclusion: He is certain that, with the prediction $\bar{P} = \bar{0}$ he will win the game if the referee measures in the diagonal basis. 

To gain information about the other outcome of the referee, Bob simulates Darwin using the information that the outcome $\bar{1}$ was communicated to him.\footnote{One may wonder why Bob does not directly predict the referee's measurement of $Q_C$. The reason is that an accurate prediction requires a description of the unitary evolution of Charly, yet Bob must consider Charly as classical to enable the previous use of \cref{assumption:consistency}.}

\noindent
\setlength{\leftskip}{1cm}
\textbf{Bob's simulation of Darwin's reasoning:} Darwin chooses his Heisenberg cut to surround $L_A Q_A L_C Q_C$ and such that Bob is classical. As Darwin has been communicated the outcome of Bob's measurement, he is certain that, Bob is certain that the outcome of his measurement is $\bar{1}$. By \cref{assumption:consistency}, Darwin is certain that Bob's measurement outcome is $\bar{1}$. The update rule then yields
\begin{equation}
  \ket{\psi'}^{\mathrm{Darwin}|Q_A = \bar{1}}_{Q_C} = \sqrt{6} \left(\leftindex_{Q_A \! \!}{\bra{\bar{1}} \ket{\psi}^{\mathrm{Bob}}}_{Q_A Q_C}\right) = \ket{1}_{Q_C}.
\end{equation}
Therefore, Darwin is certain that the referee's measurement of $Q_C$ in the computational basis will yield outcome $1$.

\smallskip
\setlength{\leftskip}{0cm}
\noindent
Bob can now again apply \cref{assumption:consistency} and, combined with his previous reasoning, conclude that he is certain that, with the pair of predictions $(P, \bar{P}) = (1, \bar{0})$, the game is won. 

Does this mean that the \ref{game:complementarity} is \textit{really} won? 
It has been shown that in quantum theory the \ref{game:complementarity} has a maximal winning probability of $\smash{\frac{1}{2} + \textstyle{\frac{1}{\sqrt{8}}}} < 1$ \cite[Eq.~(17) with $m = 2$]{Renes_2017}. The test of the referee will thus fail with non-zero probability. If it fails, we immediately have a contradiction with Bob's conclusion that the game is won with certainty. In the other case, if the game is won, this instance of the experiment does not yield a contradiction. 
However, we can provoke a contradiction with certainty by repeating the game until the physicists lose the game. According to \cref{assumption:quantum}, this will happen after finitely many repetitions. 

We conclude from this analysis that the assumptions on which Bob's reasoning were based are contradictory. We summarize this finding with the following no-go theorem.\footnote{This no-go theorem was first proposed in \cite{Frauchiger_2018}. In \cite{Frauchiger_2018}, the executability was not phrased as a separate assumption, but instead captured by making the statement conditional on the executability of the experiment. Conversely, the requirement that a physicist should not arrive at contradictory conclusions was stated explicitly as an assumption, called~(S).}
\begin{theorem}[Quantum collaboration paradox]\label{thm:FR}
  These assumptions are incompatible:
  \begin{itemize}
    \item {\textbf{Executability:}} Physicists can execute \cref{ex:consistency}.
    \item {\textbf{Universality of quantum theory:}} \Cref{assumption:quantum} holds. 
    \item {\textbf{Consistency of knowledge:}} \Cref{assumption:consistency} holds.
  \end{itemize}
\end{theorem}

Compared to the previous no-go theorems, the quantum collaboration paradox imposes a significantly stronger constraint: either we give up one of the first two assumptions, the consequences of which we already discussed after \cref{thm:obj_state}, or we give up \cref{assumption:consistency}. Contrary to the assumptions featured in \cref{thm:obj_state,thm:obj_meas} --- the state agreement assumption and the objective outcome assumption, respectively --- \cref{assumption:consistency} is operational. So, can we abolish it?

Abolishing it completely --- without replacement --- would have disastrous consequences. The reason is that it is employed extensively. For example, experimentalists gain data and communicate them to their theorist colleagues. The theorists communicate their conclusions based on this data back to the experimentalists. Without \cref{assumption:consistency}, this collaboration would not be possible.

So, does the quantum collaboration paradox mean that we need to distrust all results from experiments we did not perform ourselves? Luckily, this is not the case. At least for experiments performed today, there is a way out: All physicists simply have to agree on a common Heisenberg cut and regard themselves as one big ``meta-physicist''. However, this solution is unsatisfactory because there is no fundamental principle that would determine the location of this cut. Fixing a cut would be analogous to postulating an ether in spacetime! Indeed, as we shall see, when we move to experiments involving physicists crossing the event horizon of a black hole, a common Heisenberg cut does not generally exist. 

The quantum collaboration paradox poses problems even if there is only a single physicist performing experiments, as even then, we employ a variant of \cref{assumption:consistency}. Suppose that, yesterday, you prepared $n$ spin particles oriented in different directions, and you measure them today. To derive a statement about the time evolution of the spins, you need to combine your knowledge about their initial orientations with the knowledge of the measurement outcomes. But the former is no longer directly accessible --- at best, you find it somewhere in your lab notebook.\footnote{We assume $n$ is too large for you to remember all the spin orientations. But even if you could, your brain would just take the role of the notebook.} Thus, the knowledge about the initial spin orientations has been communicated to you over time via a physical system. Crucially, that we can use these notebook entries from yesterday in our reasoning today is an assumption --- very much in the spirit of~\cref{assumption:consistency}.

For these reasons, we need to face the challenging task of modifying \cref{assumption:quantum,assumption:consistency} in such a way that they are too weak to provoke the contradiction in \cref{ex:consistency}, but still strong enough to be usable for all practical purposes, e.g., when we communicate with other physicists. 

\Cref{assumption:quantum,assumption:consistency} have been discussed intensively in the foundations literature(see~\cite{Nurgalieva_2020} and references therein). Some argue that the assumptions are unjustified, others contend some of them as unnecessary, while still others propose specific modifications~\cite{Vilasini_2024,Renes_2021,Polychronakos_2024,DeBrota_2020,Federico_2023,Narasimhachar_2020,DiBiagio_2021}, intended to circumvent the quantum collaboration paradox. We present here a list of such arguments and then comment on them in the following paragraphs.

First, it is argued that \cref{assumption:consistency} is obviously too strong because, to accept someone else's result, one would also need to trust them to obtain the result properly and honestly report the result. 
Second, it is claimed that \cref{assumption:consistency} is unnecessary because, in all experiments performed today, the measurement result is communicated and not obtained by inference. 
Third, it is argued that \cref{assumption:consistency} can only be expected to hold for outcomes of measurements which have already been performed, and that restricting the assumption accordingly would avoid the contradiction. 
Fourth, it has been noted that, in the argument leading to the quantum collaboration paradox, one uses conclusions of physicists who will later be subject to a measurement. Hence, it is suggested that excepting these physicists from the applicability of \cref{assumption:quantum,assumption:consistency} removes the problem.\footnote{This concern has been summarized by Scott Aaronson as: ``It's hard to think when someone Hadamards your brain.'' \cite{Aaronson_2018} However, in order to avoid the contradiction of the quantum collaboration paradox, one would have to argue that ``It's even hard to think when your brain will be Hadamarded in the future.'' (See~\cite{delRio_2024} for a discussion of this issue.)} 
Fifth, it is claimed that constraining the use of \cref{assumption:quantum,assumption:consistency} to situations where no physicists are in superposition avoids any contradictions but still allows their use for all practical purposes.

To address the first argument: 
Trust in a physicist's abilities is not the point in \cref{assumption:consistency}. For Charly to be able to apply \cref{assumption:consistency}, she needs to be certain that Alice is certain about the measurement outcome and that Alice has come to this conclusion using the same theory that Charly uses. If Charly was not convinced that Alice would tell the truth or apply the theory correctly, then this would correspond to Alice applying a different theory as Charly. Hence, in this case, Charly could not apply \cref{assumption:consistency}. But this does not affect its validity.

To address the second argument:
Intuitively, it seems that communication and inference are two different concepts. 
However, what we classify as communication is also a form of inference. 
When Alice sends Bob a letter telling him the result of her measurement, then by reading the letter Bob measures it and infers that the result he read must be the result Alice got.

To address the third argument: 
Such a restriction would lead to severe practical limitations.
If an engineer is certain that tomorrow the Gotthard Base Tunnel will not collapse, then without \cref{assumption:consistency} you could not use this prediction to decide whether you risk the travel. 
It also defies the point of predictions with certainty, as the purpose of a prediction is to make a usable statement about a measurement before it happens.

To address the fourth argument: 
It is correct that a modified version of no-go \cref{thm:FR}, in which \cref{assumption:quantum,assumption:consistency} apply only to physicists who will never be measured, does not hold. However, such a modification makes the assumptions too weak to still be usable in standard situations. For example, it would limit communication, as communication necessarily counts as a measurement, thus disallowing the use of \cref{assumption:quantum,assumption:consistency}. Additionally, a physicist can usually not decide whether she is even allowed to use \cref{assumption:quantum,assumption:consistency}, as the constraint of the assumptions depends on the future. This argument is discussed in more detail in \cite{delRio_2024}.

To address the fifth argument: 
It is not clear what is meant by there being no physicist in superposition. Even in today's experiments, a physicist $\mathrm{P}_1$, who describes another physicist $\mathrm{P}_2$ together with $\mathrm{P}_2$'s environment from the outside, would typically conclude that these systems are entangled. But this means that $\mathrm{P}_1$'s description of $\mathrm{P}_2$ involves a superposition of states. Hence, a corresponding constraint on \cref{assumption:quantum,assumption:consistency} would almost always apply, preventing their use even for all practical purposes. 

\section{Black holes} \label{sec:blackholes}
The aim of this section is to argue that the findings of \cref{sec:Wigner} on Wigner's friend experiments, which did not involve any spacetime considerations, are relevant for quantum gravity. Like Wigner's friend experiments, thought experiments in gravity involve different physicists with different perspectives. To combine them, rules such as \cref{assumption:consistency} are needed. Therefore, these rules also play a role in quantum gravity puzzles.

To illustrate this, we consider specific thought experiments that closely match the Wigner's friend thought experiments described in \cref{sec:Wigner}, see \cref{tab:dictionary}. We chose these examples for concreteness, but readers who disagree with the gravity assumptions should not be deterred. We expect our central conclusion --- that rules like \cref{assumption:consistency} are essential for analysing puzzles in quantum gravity --- holds generically.

\begin{table}
    \renewcommand{\arraystretch}{1.2}
    \begin{center}
        \begin{tabular}{@{}lclcl@{}}
          \toprule
          Assumption&\phantom{-}& Wigner's friend &\phantom{-}& Black hole\\ \midrule
          State agreement && \Cref{thm:obj_state} && Hayden-Preskill experiment \cite{Hayden_2007} \\
          Objective outcomes && \Cref{thm:obj_meas} && \Cref{thm:HaydenPreskill} --- extended Hayden-Preskill \\
          \Cref{assumption:consistency} && \Cref{thm:FR} && \Cref{thm:firewall} --- firewall paradox\\
          \bottomrule
        \end{tabular}
    \end{center}
    \caption{\label{tab:dictionary} Correspondence between non-gravitational Wigner's friend and black hole thought experiments, considered in quantum foundations and quantum gravity, respectively. All these experiments depend on an assumption for how different physicists combine their perspectives. The underlying assumption is indicated in the left column. }  
\end{table}

\subsection*{Wigner's friend implies complementarity\dots{}}
To start, we consider an extension of the thought experiment proposed by Hayden and Preskill \cite{Hayden_2007}, which we describe in \cref{ex:Hayden-Preskill_v2}. As we shall see below, this protocol can be regarded as the black hole analogue of \cref{ex:Deutsch} and leads to a no-go theorem analogous to \cref{thm:obj_meas}. Furthermore, we will argue that the original protocol of Hayden and Preskill leads to the analogue of \cref{thm:obj_state}.

\begin{figure}
  \small
  \begin{tcolorbox}[colback=tintedcolor,colframe=tintedcolor]
  \begin{protocol}[Deutsch-type extension of the Hayden-Preskill experiment \cite{Hayden_2007}]\label{ex:Hayden-Preskill_v2}
    \phantom{v}

    \noindent
    \textbf{Setup:}
    \begin{itemize}
      \item Alice is freely falling into a black hole older than the Page time and carries a qubit $Q$.
      \item Bob is an outside observer of the black hole and has collected all Hawking radiation $R$ since the black hole has formed. 
    \end{itemize}

    
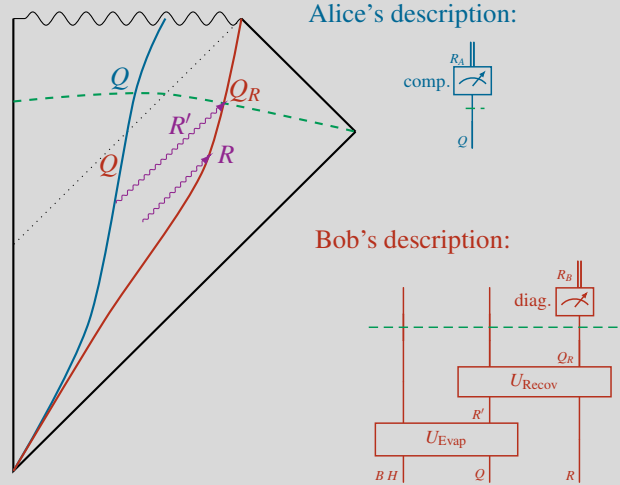
\begin{center}
      \begin{tikzpicture}
        \node[] at (0,0) {\HaydenPreskillPenrose};

        \node[MidnightBlue] at (3,3) {Alice's description:};
        \node[] at (3.5,1.8) {\HaydenPreskillAlice};
        
        \node[BrickRed] at (3,0) {Bob's description:};
        \node[] at (4,-1.9) {\HaydenPreskillBob};
      \end{tikzpicture}
    \end{center}
    
    \noindent
    \textbf{Protocol:}
    \begin{itemize}
      \item Once Alice is behind the event horizon she measures the qubit $Q$ in the computational basis and records the outcome in register $R_{A}$.
      \item Bob continues collecting Hawking radiation $R'$ until he can reconstruct $Q$. 
      \item Bob reconstructs the qubit $Q$ from the radiation $RR'$, labelling it $Q_R$.
      \item Bob measures $Q_R$ in the diagonal basis and records the outcome in $R_B$.
    \end{itemize}
  \end{protocol}
\end{tcolorbox}
\end{figure}


We analyse \cref{ex:Hayden-Preskill_v2} in the Hayden-Preskill model \cite{Hayden_2007}. In this model, for an outside physicist, the evolution of a black hole over a scrambling time corresponds to a Haar-typical unitary. It follows from quantum information theoretic arguments (decoupling theorems \cite{Dupuis_2014}) that, for an old black hole, Bob can reconstruct qubit~$Q$ from the radiation~$RR'$, as required by \cref{ex:Hayden-Preskill_v2}. 

With \cref{ex:Hayden-Preskill_v2}, Alice and Bob have achieved the same task as with \cref{ex:Deutsch}: A given qubit has been measured in two incompatible bases. Alice measured a qubit in the computational basis, Bob has reversed that measurement and measured the qubit in the diagonal basis. Another similarity to \cref{ex:Deutsch} is that no physicist can know both outcomes: Alice is in a black hole and, after Bob reconstructed $Q_R$, it is too late to jump into the black hole to meet Alice before she hits the singularity. The difference to \cref{ex:Deutsch} is that Bob does not need to isolate a laboratory but uses a black hole to ensure that he can reverse Alice's measurement. Despite this difference, it is crucial that Alice obeys \cref{principle:physical} --- in Bob's perspective, she is an ordinary physical system. In particular, her measurement is part of the evolution of the black hole, so nothing special, like a ``state collapse'', occurs. Therefore, her measurement does not inhibit Bob from reconstructing $Q$ out of the Hawking radiation.

With the same argument as in the analysis of \cref{ex:Deutsch}, i.e., employing strong subadditivity and the entropic uncertainty relation, one can show that the objectivity of measurement outcomes is in conflict with \cref{assumption:quantum} and the gravitational assumptions we have made. We phrase this conclusion as a no-go result.

\begin{theorem}[Gravitational objective outcome paradox] \label{thm:HaydenPreskill}
  These assumptions are in\-com\-patible:
  \begin{itemize}
    \item {\textbf{Executability:}} Physicists can execute \cref{ex:Hayden-Preskill_v2}.
    \item {\textbf{Universal applicability of quantum theory:}} \Cref{assumption:quantum} holds.\textsuperscript{\ref{footnote:quantum}} 
    \item {\textbf{Black hole physics:}} The Hayden-Preskill model describes a black hole accurately.\textsuperscript{\ref{footnote:page_time}} 
    \item {\textbf{Objective measurement outcomes:}} 
    A physicist's description of a physical system can be updated with all measurement outcomes ever observed.
  \end{itemize}
\end{theorem}

\refstepcounter{footnote}\footnotetext{\label{footnote:quantum} Despite the similarity of this argument and the one leading to \cref{thm:obj_meas}, there is a subtle difference in the use of \cref{assumption:quantum}. In \cref{thm:obj_meas}, \cref{assumption:quantum} could have been weakened such that a physicist can only apply quantum theory to systems she can access. Here, such a weaker variant may not apply, as it is unclear whether there is a system $S$ which both Alice and Bob can access once Bob has reconstructed $Q_R$.} 

\refstepcounter{footnote}\footnotetext{\label{footnote:page_time} This assumption must only be guaranteed until a short time after the Page time. In particular, nothing needs to be assumed about the time when the black hole reaches Planck scale.}

This theorem can be regarded as a gravitational analogue of \cref{thm:obj_meas}: both are no-go theorems questioning the assumption of objective measurement outcomes. The difference between \cref{thm:obj_meas} and \cref{thm:HaydenPreskill} is, of course, that the executability of \cref{ex:Hayden-Preskill_v2} hinges on the controllability of concrete objects like black holes instead of an abstract notion like an isolated laboratory. This is an instance of the general idea that, whenever a physicist is tasked to perform quantum operations on a macroscopic system,
he could execute this task by throwing it into a black hole and act on its Hawking radiation. 

We remark, that the analysis of the original Hayden-Preskill protocol leads to a statement analogous to \cref{thm:obj_state}. It differs from \cref{ex:Hayden-Preskill_v2} in that neither Alice nor Bob measures their qubit. Bob merely reconstructs $Q_R$ from the Hawking radiation and Alice carries the qubit $Q$. At the end of the protocol, $Q$ and its copy $Q_R$ exist on the same Cauchy slice, but they are not both accessible to a single physicist. If we nonetheless assume that there exists a joint quantum description both Alice and Bob could agree on, then the quantum no-cloning theorem is violated. To see this, consider a reference system $S$ initially maximally entangled with Alice's qubit. For Alice, this entanglement will persist, but Bob's recovered qubit would also need to be maximally entangled with $S$. This violates monogamy of entanglement. Consequently, there is no joint state of $Q Q_R S$ Alice and Bob could agree on.

One of the favoured resolutions of the contradictions arising in the Hayden-Preskill scenario is \emph{black hole complementarity} \cite{Susskind_1993,Susskind_1994,Stephens_1994}, i.e., the idea that there does not need to exist a consistent joint description among physicists who cannot communicate. In other words, the conclusion was that there is no state agreement and that measurement outcomes are not objective. These conclusions match our no-go \cref{thm:obj_state,thm:obj_meas} which arose from analysing the foundations of bare quantum theory without any gravity considerations. This is rather remarkable: thought experiments in unrelated fields --- quantum gravity and quantum foundations --- led to the same insight.\footnote{In Raphael Bousso's words: ``gravity [acted] as an oracle'' for quantum theory \cite{Bousso_2024}.}

\subsection*{\dots and that it is not enough}

To further explore the correspondence between Wigner's friend and gravity thought experiments, we proceed with the firewall paradox \cite{Mathur_2009,Almheiri_2013} and show that its conclusions can be phrased as a no-go theorem analogous to the quantum collaboration paradox. To stress the argument's operational character and similarity to \cref{ex:consistency}, we phrase the firewall paradox as a strategy to win the \ref{game:complementarity} with certainty.
\begin{figure}[h]
  \small
  \begin{tcolorbox}[colback=tintedcolor,colframe=tintedcolor]
  \begin{protocol}[The firewall strategy \cite{Mathur_2009,Almheiri_2013}]\label{ex:firewall}
    \phantom{v}

    \noindent
    \textbf{Setup:}

    \begin{itemize}
      \item Alice is freely falling into a black hole older than the Page time.
      \item Bob is an outside observer to the black hole who has collected all Hawking radiation~$R$
      emitted since the black hole has formed, and stored it in his quantum computer. For
      him, the joint state of the black hole and $R$ is pure.
      \item The referee is told to fall into the black hole and a qubit $Q_{\mathrm{Ref}}$ in the vacuum just outside the horizon~$B$ is indicated to him. 
    \end{itemize}
    \begin{center}
      \begin{tikzpicture}
        \node[] at (-1,4) {\firewallPenrose};
        \node[] at (3,1.8) {\firewallBob};
        \node[] at (4,5.8) {\firewallAlice};
        \node[MidnightBlue] at (1.4,7.3) {Alice's description:};
        \node[BrickRed] at (1.3,3.7) {Bob's description:};
      \end{tikzpicture}
    \end{center}
    
    \noindent
    \textbf{Protocol:}\footnotemark
    \begin{itemize}
      \item Alice distils a qubit $Q_{\mathrm{Alice}}$ from the modes $A$
      inside the horizon that is maximally entangled with $Q_{\mathrm{Ref}}$. 
      \item Alice measures $Q_{\mathrm{Alice}}$ in the computational basis and records the outcome $m_{\mathrm{Alice}}$ in register $R_{\mathrm{Alice}}$.
      \item Bob distils a qubit $Q_{\mathrm{Bob}}$ from the collected Hawking radiation $R$ that is maximally entangled with $Q_{\mathrm{Ref}}$.
      \item Bob measures $Q_{\mathrm{Bob}}$ in the diagonal basis and records the outcome $\bar{m}_{\mathrm{Bob}}$ in $R_{\mathrm{Bob}}$. 
      \item Bob sends $R_{\mathrm{Bob}}$ to Alice into the black hole.
      \item Alice reads the register $R_{\mathrm{Bob}}$.
      \item Alice issues the pair of predictions $(P = m_{\mathrm{Alice}}, \bar{P} = \bar{m}_{\mathrm{Bob}})$ to the referee.
    \end{itemize}
    \vspace{-0.1cm}
  \end{protocol}
\end{tcolorbox}
\end{figure}

\footnotetext{It may appear questionable if this protocol is executable. For example, Alice and Bob blatantly violate monogamy of entanglement. However, to establish such a violation, Alice's and Bob's view need to be combined, which is precisely what we want to explore.}

The concrete strategy is specified by \cref{ex:firewall}. We analyse it from Alice's viewpoint based on the following assumptions:\footnote{The statements we will derive from these assumptions are robust, i.e., it suffices that the assumptions hold approximately.}

\begin{enumerate}[label=(G${}_\arabic*$),leftmargin=.5in]
  \item\label{item:no_drama} For a freely falling physicist, all fields at the horizon are in the Minkowski vacuum state.
  \item\label{item:unitary} For an outside physicist, the evolution of a black hole and all fields (also the Hawking radiation) from the creation to the asymptotic future is a Haar-typical unitary.
  \item\label{item:effective_field} For an outside physicist, the fields outside the stretched horizon are described by effective field theory. 
\end{enumerate}

\noindent
\textbf{Alice's reasoning:}
Alice is a freely falling observer. She has access to field modes $B$ just outside and field modes $A$ just inside the horizon. By assumption~\ref{item:no_drama}, for her these modes are in a Minkowski vacuum, which is maximally entangled across the horizon. In particular, it is also maximally entangled with $Q_{\mathrm{Ref}}$. Therefore, Alice can distil a qubit $Q_{\mathrm{Alice}}$ from the vacuum just inside the horizon, which is maximally entangled with $Q_{\mathrm{Ref}}$:
\begin{equation}
  \ket{\psi}_{Q_{\mathrm{Alice}}Q_{\mathrm{Ref}}}^{\mathrm{Alice}} = \textstyle{\sqrt{\frac{1}{2}}} \left(\ket{0}_{Q_{\mathrm{Alice}}}\ket{0}_{Q_{\mathrm{Ref}}} + \ket{1}_{Q_{\mathrm{Alice}}} \ket{1}_{Q_{\mathrm{Ref}}}\right)
\end{equation}
Upon obtaining result $m_{\mathrm{Alice}}$ from her measurement of $Q_{\mathrm{Alice}}$, she can apply the state update rule~\eqref{eq:update_rule} to this state, here expressed as a circuit,
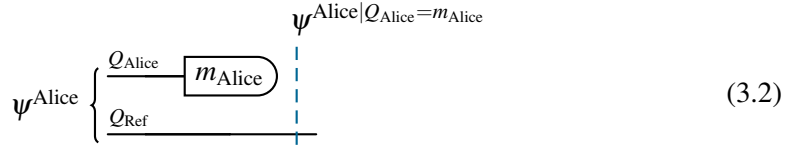
\begin{equation}
  \begin{quantikz}
    \lstick[2]{$\psi^{\mathrm{Alice}}$} & & \meterD{m_{\mathrm{Alice}}} \slice[style = MidnightBlue]{\phantom{mmmmmmmm}${\psi}^{\mathrm{Alice}|Q_{\mathrm{Alice}} = m_{\mathrm{Alice}}}$} \wire[l][1]["Q_{\mathrm{Alice}}"{above,pos=1.3}]{a}\\
    & &\wire[l][1]["Q_{\mathrm{Ref}}"{above,pos=1.2}]{a}&
  \end{quantikz}
\end{equation}

\noindent
Using \cref{assumption:quantum}, she can now predict the referee's outcome if he measures in the computational basis: she is certain that this outcome is $m_{\mathrm{Alice}}$.

Upon reading the register $R_{\mathrm{Bob}}$ she got from Bob, Alice concludes that Bob is certain his outcome is $\bar{m}_{\mathrm{Bob}}$. Using this information, she simulates Bob.

\setlength{\leftskip}{1cm}
\noindent
\textbf{Alice's simulation of Bob's reasoning:}
Bob, who is an outside physicist, describes the early radiation $R$, emitted before Alice has jumped in, and the rest of the black hole, $R'$, which still needs to evaporate. By assumption~\ref{item:effective_field}, for him, $R'$ consists of two parts: the degrees of freedom at the stretched horizon $H$ and the modes of Hawking radiation in ``the zone'' $B$.

By assumption~\ref{item:unitary}, the state of the radiation in the asymptotic future is typical. We now regard the radiation as a bipartite system consisting of the early and late radiation $RR'$. As the black hole is older than the Page time, the Hilbert space dimension of $R$ is larger than that of $R'$. Because this state is pure by the setup of the experiment, it follows from decoupling theorems that $R'$ is maximally entangled with $R$. Because $Q_{\mathrm{Ref}}$ is by definition a subsystem of $B$ and hence also of $R'$, it is maximally entangled with $R$. Therefore, Bob can distil a qubit $Q_{\mathrm{Bob}}$ from $R$ that is maximally entangled with $Q_{\mathrm{Ref}}$,
\begin{equation}
  \ket{\psi}_{Q_{\mathrm{Bob}}Q_{\mathrm{Ref}}}^{\mathrm{Bob}} = \textstyle{\sqrt{\frac{1}{2}}} \left(\ket{0}_{Q_{\mathrm{Bob}}}\ket{0}_{Q_{\mathrm{Ref}}} + \ket{1}_{Q_{\mathrm{Bob}}} \ket{1}_{Q_{\mathrm{Ref}}}\right).
\end{equation}
He measures $Q_{\mathrm{Bob}}$ in the diagonal basis and stores the outcome $\bar{m}_{\mathrm{Bob}}$ in $R_{\mathrm{Bob}}$.
Applying the state update rule~\eqref{eq:update_rule} to this state, here expressed as a circuit,
\begin{equation}
  \begin{quantikz}
    \lstick[2]{$\psi^{\mathrm{Bob}}$} & & \meterD{\bar{m}_{\mathrm{Bob}}} \slice[style = BrickRed]{\phantom{mmmmmmm}${\psi}^{\mathrm{Bob}|Q_{\mathrm{Bob}} = \bar{m}_{\mathrm{Bob}}}$} \wire[l][1]["Q_{\mathrm{Bob}}"{above,pos=1.3}]{a}\\
    & &\wire[l][1]["Q_{\mathrm{Ref}}"{above,pos=1.2}]{a}&
  \end{quantikz}
\end{equation}
Using \cref{assumption:quantum}, Bob can now predict the referee's outcome if the referee measures in the computational basis: Bob is certain that this outcome is $\bar{m}_{\mathrm{Bob}}$.

\smallskip
\setlength{\leftskip}{0cm}
\noindent
With this simulation of Bob, Alice concludes that she is certain that Bob is certain that, if the referee measures in the diagonal basis, the outcome is $\bar{m}_{\mathrm{Bob}}$. Therefore, using \cref{assumption:consistency}, she is certain that $\bar{P} =  \bar{m}_{\mathrm{Bob}}$ will win the game.

We have now reached a conclusion similar to the one for \cref{ex:consistency}: the strategy defined by \cref{ex:firewall} wins the \ref{game:complementarity} with certainty, which contradicts quantum theory.\footnote{To reach an operational contradiction, one may follow the same argument as described in the analysis of \cref{ex:consistency}. However, because the test is performed in the interior, it is more challenging to decide from the outside whether it has succeeded. But Bob may still obtain this information by reconstructing the referee from the late Hawking radiation.} In summary, our formulation of the firewall paradox as a game leads to the following no-go theorem.
\begin{theorem}[Firewall paradox]\label{thm:firewall}
  These assumptions are incompatible:
  \begin{itemize}
    \item {\textbf{Executability:}} Physicists can execute \cref{ex:firewall}.
    \item {\textbf{Universal applicability of quantum theory:}} \Cref{assumption:quantum} holds.  
    \item {\textbf{Black hole physics:}} Assumptions~\ref{item:no_drama},~\ref{item:unitary}, and~\ref{item:effective_field} hold. 
    \item {\textbf{Consistency of knowledge:}} \Cref{assumption:consistency} holds.
  \end{itemize}
\end{theorem}

One may wonder why we emphasize \cref{assumption:consistency} --- it is thought of as unproblematic, except maybe in Wigner's friend experiments. However, black holes may be used as a model of an isolated laboratory, so we actually are in a Wigner's friend situation. 

One may still ask whether the contradiction that lead to \cref{thm:firewall} can be reached without invoking  \cref{assumption:consistency}. This, however, is impossible in the above version of the thought experiment: Bob, as the outside observer, cannot describe the interior modes --- they are inside the black hole. Alice, as a freely falling observer, will end up in the black hole, so she cannot describe the late radiation $R'$. But because her prediction $\bar{P}$ is derived from Bob's analysis of $R'R$, she needs to simulate Bob and thus use \cref{assumption:consistency}.

The firewall paradox was introduced to show that, in contrast to the Xeroxing paradox, a contradiction arises even if one assumes black hole complementarity~\cite{Almheiri_2013}. Specifically, in many versions, the contradiction is constructed in such a way that it is testable by a single physicist. For example, in the version considered in~\cite{Bousso_2025}, Alice takes over the tasks Bob was supposed to do. This is possible because $Q_{\mathrm{Bob}}$ can be distilled and measured before Alice jumps into the black hole. But because assumptions~\ref{item:unitary} and~\ref{item:effective_field} require an outside perspective, their use needs to be justified. This is done by arguing that Alice, at the time when she uses these assumptions, still has the choice to stay outside. Therefore, her conclusions must agree with what she would conclude if she changed her mind to stay outside and perform a test of the assumptions (see~\cite{Bousso_2012,Harlow_2012,Bousso_2013} for the history). 

This justification implicitly uses an extra assumption (E), which ensures a statement of the following form: suppose a physicist has the possibility to perform experiment $\mathrm{Exp}_1$, but actually performs experiment $\mathrm{Exp}_2$; then the conclusions that she would have drawn, had she performed $\mathrm{Exp}_1$, also hold for $\mathrm{Exp}_2$. This assumption has a similar flavour to \cref{assumption:consistency}. But, in contrast to \cref{assumption:consistency}, which considers different physicists in the same experiment, assumption (E) considers the same physicist in different experiments. However, assumption (E) is obviously wrong.\footnote{\label{footnote:cat}Consider the following realistic example: Suppose you are in bed and want to know if your cat is in the kitchen. You know that if you go to the kitchen, you will find the cat there (expecting you to feed her). But you decide to stay in bed. Assumption~(E) now implies that the cat is in the kitchen. But, as every cat owner knows, this conclusion is false.} So if one wants to replace multiple physicists by a single physicist, one faces the challenge of finding a variant of assumption (E) that is not wrong but still allows the above conclusion.

\section{Discussion}
Non-gravitational Wigner's friend thought experiments teach us an important lesson: the universality of quantum theory, \cref{assumption:quantum}, conflicts with assumptions that are needed for combining information held by different physicists --- the state agreement assumption, the objective outcome assumption, or \cref{assumption:consistency}. This is made precise by no-go \cref{thm:obj_state,thm:obj_meas,thm:FR}.

A similar lesson was, independently, learnt from quantum gravity thought experiments. In these experiments, contradictions arise when conclusions obtained by physicists with different perspectives are combined. This was recognized and lead to the paradigm of black hole complementarity. Comparing to  Wigner's friend experiments, black hole complementarity corresponds to giving up the state agreement assumption.

In this chapter, we have deepened this correspondence. The key finding is that gravity thought experiments can be viewed as instances of Wigner's friend thought experiments, with a black hole serving as a perfectly isolated laboratory. We made this correspondence precise by relating no-go theorems originating in Wigner's friend considerations to no-go theorems concerning black holes, as summarized in \cref{tab:dictionary}. 

Our finding sheds new light on quantum gravity thought experiments. We uncover that the firewall paradox relies on \cref{assumption:consistency}, which by the quantum collaboration paradox, \cref{thm:FR}, is in  conflict with another assumption of quantum gravity, the universality of quantum theory. This opens a new avenue for resolving puzzles like the firewall paradox: resolve the quantum collaboration paradox! If successful, this would show that the firewall paradox is not genuinely rooted in gravity.

Resolving the quantum collaboration paradox means finding replacements for \cref{assumption:quantum,assumption:consistency} that are usable for all practical purposes but do not lead to a contradiction. For example, inspired by \cref{principle:physical}, \cref{assumption:quantum} could be modified such that only physicists who hold a large enough physical reference frame are eligible to apply the rules of quantum theory.\footnote{This idea has been discussed widely in the quantum foundations community but only recently been considered in quantum gravity~\cite{Gomes_2024}.} In \cref{ex:consistency}, for instance, Charly, who simulates Alice, must hold a reference for Alice. However, because Alice describes Charly, Alice needs a reference for Charly. This leads to a recursive situation: Charly's reference must serve as a reference for Alice's reference for Charly.


The correspondence between quantum foundations and quantum gravity thought experiments offers an opportunity for the two communities to learn from each other. Thought experiments in quantum gravity may inspire resolutions of the quantum collaboration paradox. For example, modelling isolated systems as black holes might reveal features of isolated laboratories not obvious in bare quantum theory. Conversely, concepts such as \cref{principle:physical}, which are crucial to understand Wigner's friend thought experiments, could be used in the study of thought experiments in quantum gravity.\footnote{Indeed, for the firewall paradox, ideas in the spirit of \cref{principle:physical} have already been used to propose a solution \cite{Yoshida_2019,Yoshida_2021}.}   

\section*{Acknowledgements}
We acknowledge support from the National Centre of Competence in Research SwissMAP and the ETH Zurich Quantum Center.

\bibliographystyle{plainurl}
\bibliography{../bibliography}

\end{document}